\documentclass{eas}
\usepackage{graphicx}
\newcommand{\msun}{\,\mbox{$\mbox{M}_{\odot}$}}

\begin{document}

\title{Measuring the Initial Mass Function of Low
    Mass Stars and Brown Dwarfs} 
\author{R. D. Jeffries}\address{Astrophysics Group, Keele University,
  Keele, Staffordshire, ST5 5BG, UK \\ rdj@astro.keele.ac.uk}

\runningtitle{Jeffries: Measuring the IMF...}
%
%
\begin{abstract}
I review efforts to determine the form and any lower limit to the
initial mass function in the Galactic disk, using observations of
low-mass stars and brown dwarfs in the field, young clusters and star
forming regions. I focus on the methodologies that have been used and
the uncertainties that exist due to observational limitations and to
systematic uncertainties in calibrations and theoretical models. I
conclude that whilst it is possible that the low-mass IMFs deduced from
the field and most young clusters are similar, there are too many
problems to be sure; there are examples of low-mass cluster IMFs that
appear to be very discrepant and the IMFs for brown dwarfs in the field
and young clusters have yet to be reconciled convincingly.
\end{abstract}
\maketitle
\section{Introduction}

The initial mass function (IMF) is the name given to the distribution of
masses with which stars and brown dwarfs are born.  Early in our
astrophysical careers we learn some form of the Russell-Vogt theorem:
that the mass and, to a lesser extent, composition of a star determine
its radius, luminosity and its subsequent evolution in the
Hertzsprung-Russell (HR) diagram. If we could determine the IMF
within a population that has broadly the same composition, then in
principle we would know almost everything about that stellar
population and its evolution.

The IMF is perhaps the most fundamental output and diagnostic of the
star formation process. As such, it acts as an external constraint that
must be satisfied by any star formation theory. We can search for
features or variations in the IMF as a function of environment or
metallicity, thus identifying the factors that are important
in shaping it (e.g. see reviews by Chabrier 2003; Bastian \etal\ 2010).

The low-mass IMF may be especially important as a star formation
diagnostic. There is abundant evidence that low-mass stars
($m<0.3\,M_{\odot}$) and brown dwarfs (BDs, $m<0.075\,M_{\odot}$) form
in a similar way to stars like the sun (e.g. Luhman \etal\ 2007;
Whitworth \etal\ 2007, 2010), but in many cases they must have masses
well below any average thermal Jeans mass of the cloud where they were
born. So what produces them and what stops them from accreting more
mass? Many ideas have been put forward (ejection, turbulent
fragmentation, photoevaporation) and are discussed elsewhere in this
volume, but the important point is that these processes will put their
imprint on the IMF. Some excellent examples of this can be found in a
series of papers where smoothed particle hydrodynamic simulations are
used to investigate how the inclusion of different pieces of physics,
or differing intial conditions, can affect the IMF (Bate \& Bonnell
2005; Bate 2005, 2009a, 2009b).

The IMF also provides a strong link between stellar physics and the
physics of galaxies. Bastian \etal\ (2010) introduce (tongue-in-cheek)
Dav\'e's theorem; that all problems in extragalactic astrophysics can
be solved with an appropriate choice of IMF! It is certainly true that
a knowledge of the IMF is vital in understanding the observable
properties and mass budget of whole stellar systems and unresolved
populations.  For typical assumptions for the form of the IMF, more
than 60 per cent of the stellar mass is in relatively dim stars with
$m<1\,M_{\odot}$. Some of the highest profile results in extragalactic
astronomy, such as the Schmidt-Kennicutt law relating galactic star
formation rates to the surface density of gas (Schmidt 1959; Kennicutt
1998), or the ``Madau diagram'' plotting the star-forming rate in the
universe as a function of redshift (Madau \etal\ 1996, 1998; Gonzalez
\etal\ 2010), rely to some extent on an assumed IMF. The observational
tracers of star formation activity (H$\alpha$, rest-frame UV continuum
etc.) generally trace luminous high-mass populations that only make up
a small fraction of the total stellar mass. Mass-to-light ratios and
star formation rates, in solar masses per year, are inferred from
synthesis models (e.g. Bruzual \& Charlot 2003) combined with
assumptions about the IMF.

In this article I summarise a series of lectures that were given as
part of the 2011 Evry-Schatzman school on ``Low-mass stars and the
transition from stars to brown dwarfs''. The objective of these
lectures was to review progress towards measuring the IMF of the
Galactic disk based on observations of low-mass stars and brown dwarfs
in the field and in young clusters. There is an emphasis on examining
the methodologies used in these scenarios and I focus on weaknesses in
techniques or uncertainties in physical models that render some results
more reliable than others.

In section~\ref{lec1} I start with a historical perspective and examine
how (sub)stellar masses are deduced from observational indicators and
how mass-luminosity relations are calibrated. In section~\ref{lec2} I
examine attempts to determine the low-mass stellar IMF from a census of
local field stars, and this is extended to a consideration of the
substellar field IMF in section~\ref{lec3}, where different techniques
need to be used. Section~\ref{lec4} is devoted to measuring the IMF in
open clusters and star forming regions and the problems that must be
overcome to yield reliable results. Section~\ref{lec5} reviews attempts
to find any lower limit to the IMF in clusters and the field.
Section~\ref{summary} provides a summary of the major points, discusses
areas where progress needs to be made and perspectives for future
research.

\section{Luminosity Functions to Mass Functions}

\label{lec1}

\subsection{A historical perspective}

The history of IMF and MF determinations begins with studies of the
luminosity functions (LFs) of Galactic field populations conducted in
the early part of the 20th century (read the excellent review of Reid
\& Hawley 2000). Perhaps the main motivation at that time for
determining $\Phi (M)$, the number of stars per unit absolute magnitude
($M$) interval per cubic parsec, was to interpret star count data in an
effort to understand the structure of our own Galaxy.

Early attempts to determine $\Phi(M)$ were of course hampered by a lack
of accurate trigonometric parallaxes. A number of clever statistical
techniques -- the method of mean parallaxes or the use of relationships
between absolute magnitude and ``reduced proper motion'' -- were
devised and exploited by pioneers such as Kapteyn and Luyten (see for
example Kapteyn 1902; Luyten 1923). There were also significant
problems with incompleteness for early samples selected from their
large proper-motions. Later studies such as those of Kuiper (1942)
attempted to construct volume-limited samples of nearby stars using
trigonometric parallaxes, supplemented by stars with distances
estimated from a relationship between absolute magnitude and colour --
a photometric parallax. There were (and still are) significant
uncertainties associated with such relationships for cool, low-mass
stars. This was partly because the colour commonly used at that time
was $B-V$, which becomes degenerate for $T_{\rm eff}\leq 3600$\,K, but
also because few stars at these temperatures were bright enough to have
measured parallaxes that could be used for calibration.  As a
consequence, very little could be said for certain about the LF for
$M_V>11$ (spectral types cooler than about M2).

The situation has of course improved a great deal in the last few
decades and will be discussed in more detail in section~\ref{lec2}, but
two clear points emerge from these early studies. (i) That
incompleteness in samples of nearby stars is often a major problem when
trying to study the IMF at low-masses and (ii) that even should the
determination of $\Phi(M)$ be exquisitely accurate, we still need a way
of converting luminosities to masses before the MF or even the IMF can
be estimated.

\subsubsection{Some definitions and formalism}

Much of the formalism adopted in IMF studies is due to
Salpeter (1955) and Miller
\& Scalo (1979), who looked at the LF of field stars
in the Galactic disk and
made attempts to determine the IMF from it. First there are the
definitions of the luminosity and mass functions.
\begin{equation}
\Phi (M) = \frac{dN}{dM}
\end{equation}
is the number of stars per unit magnitude ($M$) interval, per cubic
parsec.
\begin{equation}
\phi( \log m) = \frac{dN}{d \log m}
\end{equation}
is the number of stars per unit logarithmic mass ($m$) interval, per
cubic parsec. Sometimes a linear mass function ($\phi (m) = dN/dm$) is
preferred, but given that the IMF covers 4 orders of magnitude (from
high mass stars to planets) and there are some arguments that the IMF
may be log-normal in form (see below), then the logarithmic representation is
commonly adopted.

There are a number of ways that the MF or IMF can be parameterised. The
most common are a power law form with index $\alpha$
\begin{equation}
\phi(m) \propto m^{-\alpha} \ \ \ {\rm or} \ \ \ \phi(\log m) \propto
m^{-\alpha + 1}\, ,
\label{powerlaw}
\end{equation}
or the log-normal form
\begin{equation}
\phi(\log m) \propto   \exp \left(-\frac{(\log m - \log m_c)^2}{2
  \sigma^2}\right)\, ,
\label{lognorm}
\end{equation}
that has a width parameter $\sigma$ and a characteristic mass $m_c$
(Chabrier 2002, 2003, 2005). Other more exotic forms have been
proposed, for example a smoothed two-power law
\begin{equation}
\phi(m) \propto m^{-\Gamma} (1 - \exp[-(m/m_p)^{\gamma + \Gamma}])
\end{equation}
that has power law indices $\Gamma$ at high masses, $\gamma$ at
low-masses and a transition mass $m_p$ (de Marchi \& Paresce 2001;
Parravano \etal\ 2011). There is also the simple expedient of modelling
the MF as a series of connected power law relationships with
characteristic transition masses (e.g. Kroupa 2002).

If an attempt is made to measure the MF from field stars, this is {\it
not} the IMF, because massive stars with short lives may well have
completed their evolution and become white dwarfs, neutron stars or
black holes, and will hence be under-represented. Some fraction (that
increases with mass) of all stars with $m\geq 0.9\,M_{\odot}$ will have
evolved from the main-sequence given an age for the Galactic disk
$T_{\rm G} \simeq 10$\,Gyr. We refer to the MF measured in this way as
the {\it present day} mass function (PDMF, $\phi_{\rm PD} (\log
m)$). Miller \& Scalo (1979) defined a ``creation function'' $C(\log m,
t)$, which is the rate at which stars form per unit (logarithmic) mass
interval per cubic parsec. If we assume that the IMF is constant in
time, then the mass dependence and time dependence are separable and
allow the PDMF to be written as
\begin{eqnarray}
\phi_{\rm PD} (\log m) & = &   \phi (\log m)\, \int^{T_{\rm G}}_{T_{\rm
    G}-T_{\rm MS}} f(t)\ dt \ \ \ \ {\rm for}\ T_{\rm MS}<T_{\rm G}\, ,
\\
\phi_{\rm PD}(\log m) & = & \phi(\log m) \ \ \ \ \ {\rm for}\ T_{\rm
  MS}\geq T_{\rm G}\, ,
\end{eqnarray}
where $T_{\rm MS}$ is the main sequence lifetime for a star of mass
$m$, and $f(t)$ is an arbitrary function of time that encapsulates the
star-forming history of the Galactic disk. $f(t)$ is normalised such
that its integral over time from zero to $T_{\rm G}$ is unity. Thus we
see that for stars with $M>0.9\,M_{\odot}$, some estimate of the star
forming history is required to transform the PDMF to the IMF. For less
massive stars the PDMF {\it is} the IMF, but we shall see that there is
an additional complication that means that $f(t)$ becomes important
once more for stars and BDs with $m\leq 0.1\,M_{\odot}$.

\subsubsection{Early attempts to obtain the IMF}

The basic observational experiment is to determine the LF in the form
$dN/dM$ and then multiply this by the slope of a mass-magnitude
relationship, $dM/d \log m$, to get $\phi_{\rm PD}$. An assumed star
forming history is then used to correct $\phi_{\rm PD}$ and obtain the
IMF. Taking an LF determined by van Rhijn (1936) and Luyten (1941), a
crude mass-magnitude relationship and assuming a uniform star forming
history, Salpeter (1955) was able to propose that $\phi(m) \propto
m^{-\alpha}$, with $\alpha = 2.35$, for stars with $0.4 < m/M_{\odot} <
10$. Miller \& Scalo (1979) investigated various star forming histories
and updated the field LF using what were the best proper-motion and
parallax surveys of the time (e.g. Luyten 1968; Wielen 1974). Miller \&
Scalo found their IMF was best represented by a semi log-normal
distribution with a characteristic mass of $0.1\,M_{\odot}$ and width
of 0.7 dex for stars with $0.1 < m/M_{\odot} < 50$. This
parameterisation roughly agrees with Salpeter over the mass range
common to both, but corresponds to a larger value of $\alpha$ at
$m>10\,M_{\odot}$ and smaller $\alpha$ for $m<0.4\,M_{\odot}$.

Both of these highly cited early attempts to measure the IMF are
entirely reliant on the fidelity and completeness of the low-mass star
surveys and an accurate relaionship between mass and magnitude. The
former will be discussed in section~\ref{lec2}, the latter is discussed here.

\subsection{Mass-Magnitude relationships for low-mass stars}

The conversion of a LF to a MF relies on a mass-magnitude
relationship. This can be taken from evolutionary models of stars,
especially for pre-main sequence (PMS) or very low-mass objects (see
sections~\ref{lec3} and~\ref{lec4}), but can be determined empirically
for low-mass main sequence
stars. The requirements for this
calibration are stars with a a range of masses that have an accurate
distance and photometry, in order to determine their {\em absolute}
magnitude, and of course a means of measuring their mass. The former is
most usually found through a trignometric parallax, but the latter is
more difficult to obtain.

\subsubsection{Stellar masses}
The masses of stars come chiefly from measurements of stars in binary
systems -- either resolved, astrometric binaries or close, eclipsing binaries.

In an astrometric binary, the relative orbit is an ellipse with the
primary star at one focus. An inclined orbital plane means that the
projected orbit on the plane of the sky is also an ellipse, but
the primary is not at the focus. The displacement of the primary from the
focus yields the orbital inclination and then the semi-major axis and
eccentricity can be deduced. The semi-major axis and orbital period
give the {\it total} system mass from Kepler's third law, but the
semi-major axis needs to be in physical, rather than angular units, so
the distance to the binary is required (see Kraus \etal\ 2009 for an
example of this approach). 
If one has {\it absolute} orbits (i.e. the RA and Dec of each
component as a function of time, rather than just separation and
position angle), then the ratio of the apparent orbital displacements
gives the inverse mass-ratio. Combined with the system mass this gives
the masses of the individual components. Likewise, the mass-ratio can
be constrained by measurements of the radial velocity (RV) curves
of the two components (e.g. see Andersen \etal\ 1991; S\'egransan \etal\
2000 for applications to low-mass stars).

In an eclipsing binary, the mass-ratio again comes directly from the
amplitude ratio of the RV curves and gives one constraint on the
individual masses. The system inclination (and the
stellar radii) can be obtained by modelling the eclipses and then a
second constraint comes from the dynamical mass function
\begin{equation}
\frac {m_{2}^{3} \sin^3 i}{(m_1 + m_2)^2} = \frac{P K_{1}^{3}}{2\pi
    G}\, ,
\end{equation}
where $P$ is the binary orbital period and $K_1$ is the measured RV
amplitude of the primary. This leads to
component masses without the need for an accurate parallax
(e.g. see Morales \etal\ 2009; Torres \etal\ 2010).

\subsubsection{The empirical and theoretical  mass-magnitude relationships}

\label{massmag}

\begin{figure}

\begin{center}
\includegraphics[width=10cm]{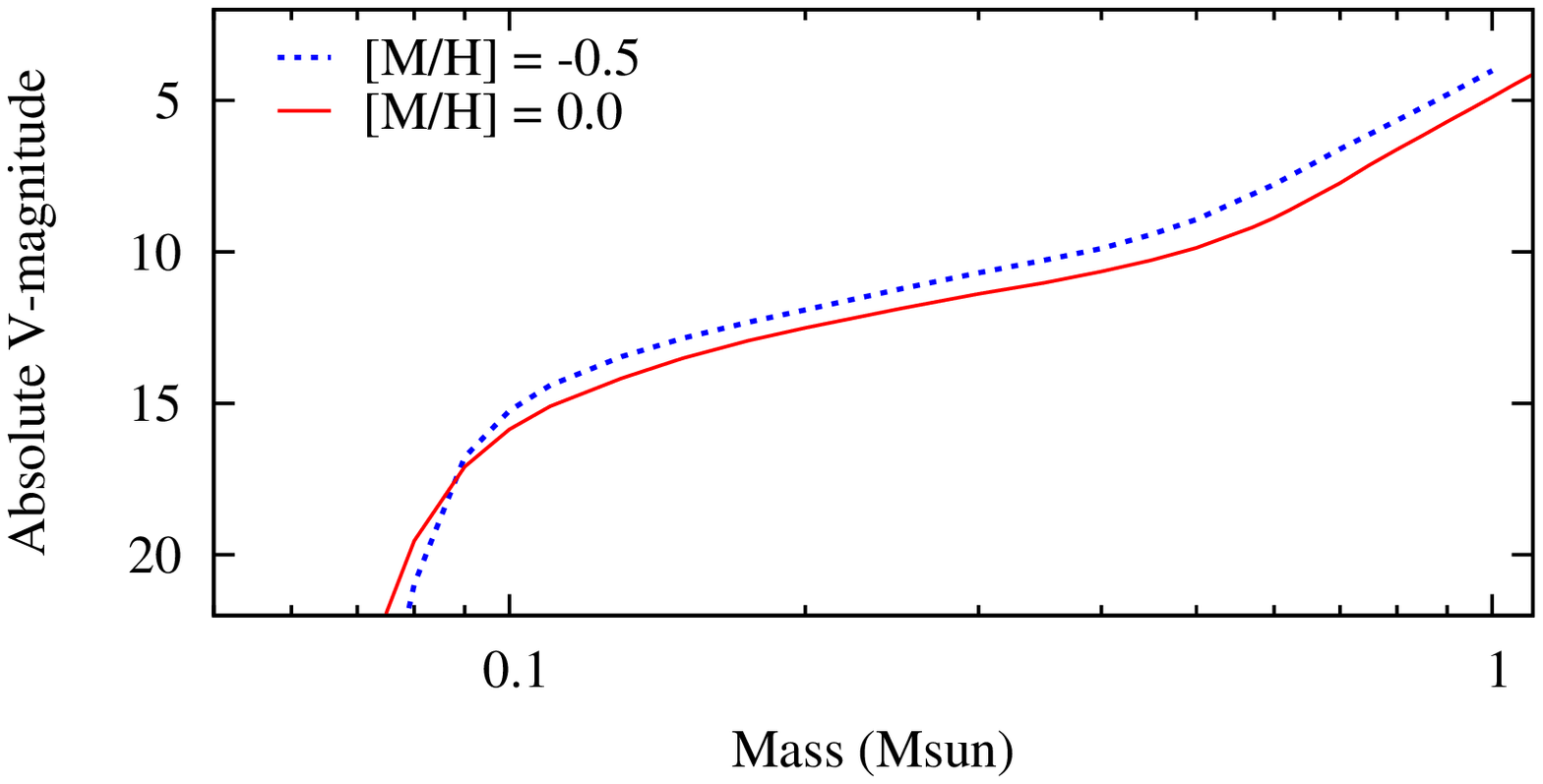}
\includegraphics[width=10cm]{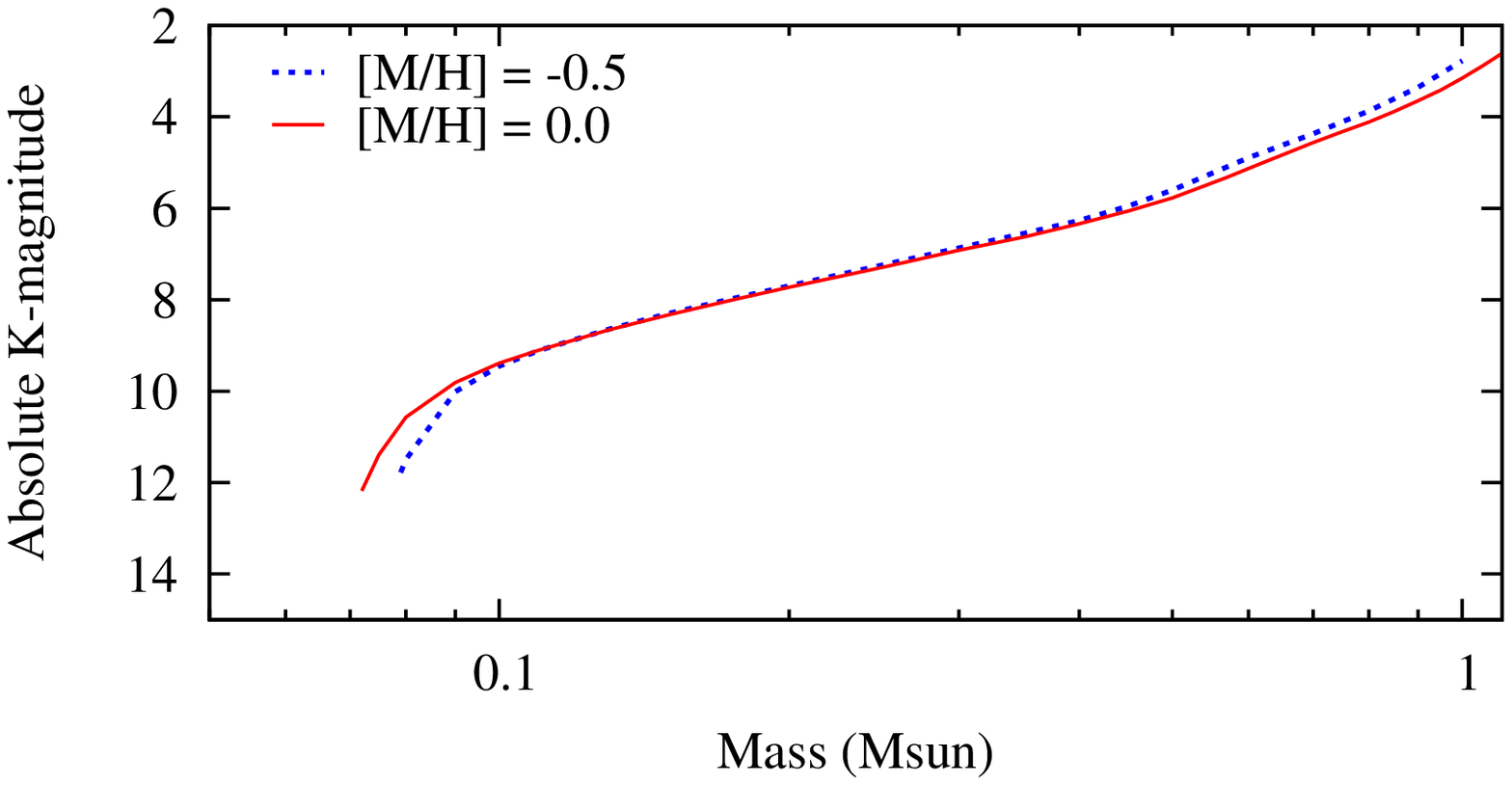}
\end{center}

\caption{Theoretical mass-magnitude relations (for an age of 5\,Gyr)
  for the $V$ and $K$ bands,
  taken from the models of Baraffe \etal\ (1998). These show that
  masses estimated from absolute $V$ magnitudes are more susceptible to
  systematic errors arising from metallicity uncertainties.}

\label{massmagvsz}

\end{figure}

\begin{figure}
\begin{center}
\includegraphics[width=10cm]{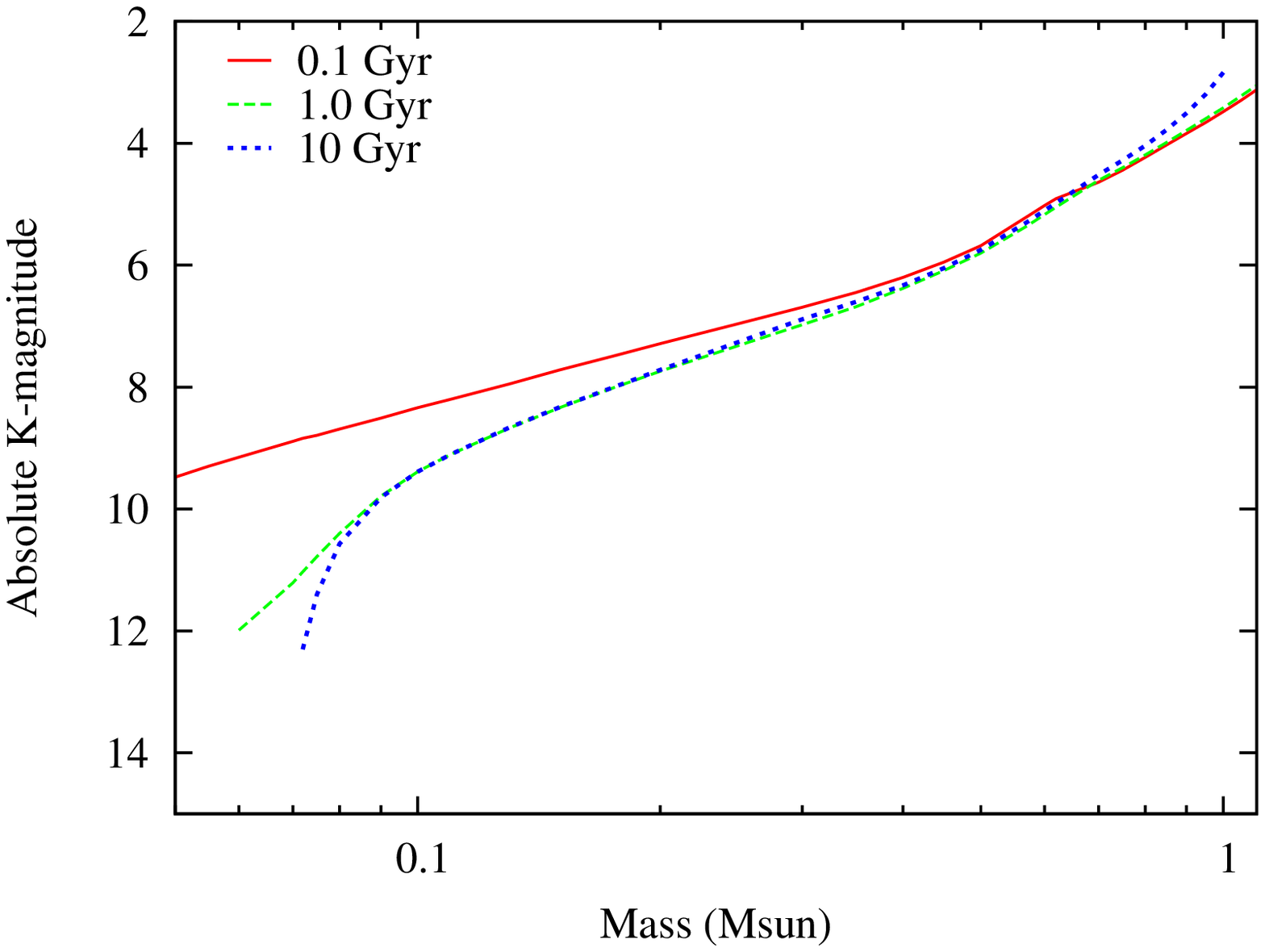}
\end{center}
\caption{The age dependence of the $K$-band mass-magnitude relation,
  taken from the models of Baraffe \etal\ (1998).}
\label{massmagvsage}
\end{figure}

After measuring masses, the photometry and parallaxes are combined to
give an empirical mass-magnitude relationship. This can be used
straightforwardly to derive mass functions from luminosity functions. A
comparison with models is valuable though, because we do need to
understand some details of the structure and atmospheres of low-mass
objects in order to identify possible sources of systematic error in
the relationships. It is also a pre-requisite that models match the
data we have before we can have any confidence in extrapolating these
models to ages and masses that are not yet well-calibrated by
observational data.

The best-known relationships for low-mass stars are those presented in
Delfosse \etal\ (2000, see their Fig.3; but see also Henry \& McCarthy
1993; Xia \etal\ 2008). The main conclusion that can be drawn from this
study is that the tightest relationships are found between
near-infrared absolute magnitudes and mass. These are in good agreement
with the main-sequence properties predicted by evolutionary models such
as those of Baraffe \etal\ (1998) and Siess \etal\ (2000) over the
range $0.1<m/\msun < 0.8$.  In contrast there is much more scatter
(about one magnitude, or about 30 per cent in mass) in the $V$-band
relationship, and the models show significant discrepancies with data
at the lowest masses.

The likely explanation is that the $V$-band magnitudes of cool,
low-mass stars are highly dependent on some uncertain atmospheric
opacities, that themselves are quite metallicity dependent (see
elsewhere in this volume and Fig.~\ref{massmagvsz}). A field population
of low-mass M-dwarfs will have a range of metallicity, so the scatter
in the mass-magnitude relation is not unexpected, but the overall
discrepancies probably indicate remaining problems in calculations of
atmospheric opacities at low temperatures. The near-infrared magnitudes
are much less uncertain and much less dependent on metallicity. The
lesson is clear -- the most reliable results will come from
near-infrared assessments of field luminosity functions.

A further issue to consider is any age dependence of the mass-magnitude
relation. At $m>0.8 \msun$ the oldest field stars will have started to
increase in luminosity as they evolve on the main sequence (see
Fig~\ref{massmagvsage}) and this must be accounted for in any detailed estimate of
the MF. At younger ages, low-mass objects are still descending
towards the ZAMS (which they will never reach if
$m<0.07\msun$). The PMS timescale is roughly equivalent to the
Kelvin-Helmholtz time, which is approximately proportional to $m^{-1}$.
For a uniform star formation rate over 10\,Gyr, about 10 per cent of
stars with $m=0.1\msun$ will still be on the PMS, thus estimates of the
field MF could safely use a main sequence mass-magnitude relationship
for $0.1<m/\msun <0.8$ (see Fig.~\ref{massmagvsage}), 
but stars outside of this range would be more
luminous (on average) than a main sequence relationship would
predict. Of course this time-dependence can be important at all masses when
dealing with young clusters (see section~\ref{lec4}). 

\subsection{Mass-Magnitude Relationships for Brown Dwarfs}
\label{bdml}

For $m\geq 0.08\msun$ the core temperature eventually becomes high enough to
initiate hydrogen fusion, but the cores of objects with $m<0.07\msun$
never attain this status; such objects 
reach a minimum radius and then simply cool and fade
as degenerate BDs (see Chabrier \& Baraffe 1997; Burrows \etal\
1997). A further complication is that the brief phase of deuterium
burning in the early PMS is more extended in the lowest mass objects,
perhaps 100\,Myr at $m=0.015\msun$, but at even lower masses, even
deuterium burning does not occur.

This complicated luminosity evolution means that BDs are never
stationary in luminosity or the HR diagram. This offers problems and
opportunities. In principle, like PMS stars, position in the HR diagram
could determine both mass and age (if the models were correct!). In
practice, a typical $T_{\rm eff}$ measurement uncertainty (for instance
from a spectral type) of $\pm 150$\,K, which may also have systematic
errors, encapsulates BD ``cooling tracks'' with a mass range of
factors of 2--3 (see Fig.~11 in Burrows \etal\ 1997). Furthermore,
although the various available models are in reasonable agreement on
the mass-luminosity relation as a function of time, they do  not agree
on the $T_{\rm eff}$ scale, so that different models would lead to very
large differences in masses deduced from the HR diagram (Konopacky
\etal\ 2010).

Progress can be made by testing the theoretical mass-luminosity
relationships for very low mass objects. If the model
predictions could be confirmed for a few benchmark objects then we
could be more confident in extrapolation or interpolation to other
masses and ages.

Dynamical masses for BDs can be estimated in similar ways to those
described earlier for low-mass stars. However, Kepler's third law tells
us that $(m_1 + m_2) \propto a^3 /P^2$, where $a$ is the semi-major
axis. To measure orbital motion on decadal timescales requires
separations of less than a few au. BD binaries, even at 10\,pc, 
would still be very faint and have angular component separations less than a
few tenths of an arcsecond. Separating these requires high
spatial resolution that until recently could only be achieved for a few
nearby objects with adaptive optics (Close \etal\ 2003) or from space
(Bouy \etal\ 2003).

The 2MASS, SDSS and UKIDSS surveys have resulted in an explosion of BD
discoveries, approximately 10 per cent of which are in binary systems. 
Most of these are not bright enough to act as their own natural guide stars,
so it is only with the advent of laser guide star technology that
large programs monitoring many potential binary systems have become
possible (Dupuy \etal\ 2009; Konopacky \etal\ 2010).

One of the first BD masses to be confirmed was GL\,569Bab with a system mass
of $0.123\msun$ (Lane \etal\ 2001). This 890-day binary has now been
followed for more than a decade, and RV measurements give component
mass of 0.073\msun\ and 0.053\msun\ (Konopacky \etal\ 2010). Compared with
the theoretical HR diagram and the appropriate cooling tracks it is
found that whatever its age and whichever models are used, 
the GL569\,Bab system is either too luminous
or the temperatures determined from matching synthetic models to near
infrared spectra are too cool. To put it another way, the mass deduced
from the HR diagram would be far too low.
Konopacky \etal\ (2010) have conducted 
this kind of analysis for a dozen low mass L-dwarf binaries and found a
consistent picture of dynamical masses that are higher than
expected. This could be a problem either with the evolutionary models
or perhaps it is a problem with the atmospheres used to model the
effective temperature. It is also of interest that the one T-dwarf
binary studied by Konopacky \etal\ has a lower dynamical mass than expected.

A different kind of test and one which is more incisive for IMF studies
is to investigate the theoretical mass-luminosity relationship
for binaries with known age. Unfortunately, BD binaries in the nearest
open clusters are currently beyond reach, but more indirect methods of
estimating age can be used. Close \etal\ (2005) used adaptive optics to
measure the orbit and estimate the mass of AB Dor C, a faint companion
to a young star which is the prototype member of a proposed coeval
kinematic group in the solar neighbourhood. For a mass of 0.09\msun,
and an assumed age for the AB Dor association of 50\,Myr, Close \etal\
concluded that AB Dor C is a factor of 2--3 underluminous compared with
model predictions. However, the ages of these kinematic groups are not
beyond dispute. Luhman \etal\ (2005) reassessed the age of the AB Dor
association, finding an age of 100--125\,Myr. At this older age, the
models predict the {\it correct} luminosity for AB Dor C.

Another recent example is HD\,130948BC, a close-to-equal luminosity
(and hence mass) L-dwarf binary around a field G-star with a known
rotation period and metallicity. Dupuy \etal\ (2009) obtained a relative
orbit and system mass of 0.109\msun. The rotation of the G-star
suggests an age of 800\,Myr.  Like other BD binaries (see above), the
components of HD\,130948BC are either too luminous or too cool compared
with the theoretical HR diagram cooling tracks. The relatively precise
age determination ($\pm \simeq 200$\,Myr) shows that using the measured
luminosity and current evolutionary models would over-estimate the
component masses by 30 per cent.

A contrasting picture is found for the recently discovered system
$\epsilon$ Indi Bab.  This has great potential to constrain models,
because its components are very cool. It is a T1+T6 close binary in a
wide orbit around a K4.5V star at 3.6\,pc. The 12 year T-dwarf binary
orbit gives a dynamical system mass of 121\,$M_{\rm Jup}$ (Cardoso
\etal\ 2010), which given their luminosities indicates that both
components are likely to be BDs. The positions of the components in
colour-magnitude diagrams are not well described by current models
(King \etal\ 2010). This is not a new result (see later), but what is
more alarming is that while the magnetic activity of the K4.5V primary
star indicates an age of $<2$\,Gyr, the measured luminosity and
dynamical mass demand an age of $4\pm 0.3$\,Gyr. In other words, just
using the measured luminosity and the age from magnetic activity would
result in the deduction of a mass that was {\it at least} 25 per cent
{\it too low}.

In summary, whilst the mass-magnitude relationships for low mass stars
with $m>0.1\msun$ are reasonably secure, there remain significant
uncertainties when converting luminosities to masses in the case of
objects with $m<0.1\msun$ and a hint of discrepancies that may vary
with the age or surface temperature of the object.  In addition to the
physical uncertainties introduced by an age dependence in the
mass-luminosity relation and the difficulty of estimating the ages
unless objects are in a coeval cluster, it seems quite likely that
there are some quite important systematic errors in the evolutionary
models and/or the model atmospheres that remain to be resolved.

\section{The Initial Mass Function from Low Mass Field Stars}

\label{lec2}

In this section I discuss measurements of the Galactic disk IMF using
observations of low-mass field stars. The PDMF is roughly equivalent to
the IMF for stars with $0.1<m/\msun <0.8$, in the sense that the
effects of stellar evolution on the mass-magnitude relations are
minimal (section 2.1.1).  Of course the PDMF of an arbitrary star
sample might still not be the IMF if there were any significant
variation of the IMF in time or space.

The basic technique consists of three steps: (i) Measure the luminosity
or absolute magnitude of stars that lie within some defined
volume; i.e. compute the LF. (ii) Convert the luminosities or absolute
magnitude to masses using the relationships discussed in the last
section. (iii) Correct the derived MF (or the intermediate LF) for
incompleteness, biases, binarity, Galactic structure and stellar
evolution. 
It is also possible to use a more inductive approach. This would
assume a form for the IMF to predict the star counts or apparent
magnitude distribution of a sample by folding it through a
mass-magnitude relation, a Galactic structure model and then filter the result
according to observational sensitivity and
uncertainties. However, the former three-point procedure is
pedagogically more sensible to describe here and within that scheme
there are broadly two routes.

The first uses star counts in the solar neighbourhood to build up a
volume limited sample. An implicit assumption is that the gradual
dispersion of Galactic orbits means that the solar neighbourhood offers
a representative sample of stars from the disk, even if
modest corrections to the IMF are needed to account for the Sun's present
position. The advantages of this technique arise from the
proximity of the stars. Even very low-mass stars are reasonably bright
if nearby; distances can be measured with potentially very precise
trigonometric parallaxes and it is possible to resolve a large fraction
of any binary population. On the other hand the sample may suffer
from uncertain levels of incompleteness because of the inhomogeneous
methods used to identify nearby stars. Samples may also have 
small number statistics if
selected from a very small local volume, and could be afflicted by
Lutz-Kelker bias (see section~\ref{lutzkelker}).

A second method can be characterised as a photometric field star
survey. These may be wide and shallow or very deep pencil-beam
surveys. The advantages here are that one can obtain very large samples
and the level of incompleteness can be well controlled. Against this
must be set the disadvantages of relying on photometric parallaxes for
distance measurements, Malmquist bias, unresolved binarity and a much
greater reliance on Galactic structure models in the interpretation of
deep surveys.

\subsection{The IMF from the local field population}

\subsubsection{A census of the solar neighbourhood}

Since the beginning of the 20th century proper
motion measurements have been used to identify nearby stars. Typical tangential
velocities combined with proximity leads to high proper motion  
(e.g. Luyten 1924; van Maanen 1937). The early work culminated in
classic works collating high proper motion stars -- the Luyten
Two-Tenths catalogue (LTT -- Luyten 1957) the
Luyten Half Second catalogue (LHS -- Luyten 1979a) and the New Luyten
Two Tenths catalogue (NLTT - Luyten 1979b). These surveys, which
subsequently formed the basis for most work on nearby stars, are
strongly biased against the inclusion of southern hemisphere stars, and
to this day there are still possibilities to find new, relatively
bright, nearby stars in the south (e.g. Hambly \etal\ 2004; L\'epine 2005).

Proper motions were used by Gliese (and later Jahreiss) to
compile the first, second and third catalogues of nearby stars, within
20\,pc, 22\,pc and 25\,pc respectively 
(Gliese 1957; Gliese 1969; Gliese \& Jahreiss 1991 -- CNS3). 
About half of the stars in this latter catalogue 
had trigonometric parallaxes; the rest
were included on the basis of photometric and spectroscopic parallaxes
(many have since proved to be erroneously included). In the 1990s the
Palomar Michigan State University (PMSU) survey conducted extensive
spectroscopic investigations of CNS3, refining it and defining
complete samples to magnitude-dependent distance limits (Reid \etal\
1995; Hawley \etal\ 1996; Gizis \etal\ 2002). Throughout the last
decade efforts have focused on (i) pushing these limits further outward for
faint low-mass stars using the NLTT/2MASS catalogues and spectroscopy
(Reid \etal\ 2004; Cruz \etal\ 2007); (ii) using reanalyses of older plate
material to increase completeness (L\'epine \etal\ 2003; L\'epine 2005);
and (iii) establishing better trigonometric parallax data for the very
nearest objects (Henry \etal\ 2006). It is now likely that {\it for
  stars}, the sample is complete to 8\,pc in the northern
hemisphere, with further work to do in the south; is rapidly becoming
complete to 10\,pc; and is about 80 per cent complete to 20\,pc (Reid \etal\ 2007). 

\subsubsection{Lutz-Kelker bias}

\label{lutzkelker}

How is a nearby sample constructed? Unfortunately, even if
trigonometric parallaxes are available, there are significant problems
in using these unless they are precise. Lutz \& Kelker (1973)
showed that if stars are uniformly distributed in
space then more are scattered {\it in} to a sample from a larger true
parallax than are scattered out. This means that if the true parallax
is $\pi_0$, the observed parallax is $\pi$ and the uncertainty in the
measurement is $\sigma_\pi$, then the mean value of $\pi_0/\pi$ is less
than unity. The estimated absolute magnitude of such a sample will be
biased such that $M_{\rm true} = M_{\rm obs} + \Delta M$. The bias $\Delta
M$ is negative and a rapidly increasing function of $\sigma_\pi /\pi$, such
that $\Delta M = -0.11$ mag for $\sigma_\pi /\pi=0.1$, $\Delta M =
-0.28$ mag for $\sigma_\pi /\pi=0.15$ and essentially no meaningful
correction can be made for $\sigma_\pi /\pi \geq 0.2$. In other
words, contamination of the sample with objects that are at up to an 
infinite distance becomes important! In practice, the corrections may be
complicated by the fact that the parent sample is not uniform, for
example there could be growing incompleteness at larger distances
(Hanson 1979).

The Lutz-Kelker 
effect was amply demonstrated when some of the earlier nearby star
catalogues were compared with much more accurate parallax data (for
bright stars) from the Hipparcos satellite (Jahreiss \& Wielen
1997). About one third of the original CNS3 sample was excluded, and this
fraction would have been higher if many stars had not already been
excluded from CNS3 using additional photometric/spectroscopic parallax
information. 

A further important application that may be susceptible to this bias is
when samples with trigonometric parallax are used to calibrate
mass versus absolute magnitude or absolute magnitude versus colour
relationships. Such relationships should be treated with extreme caution if the
samples used have parallax uncertainties any larger than 10 per cent.
  
\subsubsection{A case study}

A representative (and pedagogic) example of the nearby star approach is
provided by Reid \etal\ (2002). These authors constructed a volume
limited sample for $8<M_V < 16$ (roughly spectral types M0V and
cooler), confined to declinations
$> -30^{\circ}$ (558 stars in 448 systems). This was supplemented with
Hipparcos data for $M_V<8$ (1028 stars in 764 systems), which after filtering
out giants and white dwarfs, should be complete to 25\,pc.
About 90 per cent of this sample have precise trigonometric parallaxes.

A number of steps were taken to assess the {\it completeness} of the
sample. There is a trivial spatial incompleteness in the faint sample
that is easily dealt with. Another factor considered and rejected was
whether there was a bias away from finding high proper motion stars in
crowded fields close to the Galactic plane, though there was no
investigation of this as a function of magnitude. 
{\it Differential} completeness among fainter stars 
was assessed in absolute magnitude bins by plotting stellar density as a
function of distance, under the assumption that it should be
uniform. This resulted in a completeness limit of 20--22\, pc for stars
with $M_V \simeq 9$, but only 5\,pc at $M_V =15.5$.

Two futher important issues were investigated. The first was
whether selection by proper-motion leads to incompleteness. As the
nearby stars were predominantly identified on the basis of their large
proper motions it is possible that objects with low proper motions,
which are more likely to be at larger distances, have not been
identified. For instance, detection in the NLTT catalogue at a proper
motion of $>0.2$ arcseconds/yr corresponds to a tangential velocity of
$>20$\,km\,s$^{-1}$ at 20\,pc, which some stars may not possess.  This was tested by
comparing the transverse motions with those of a sample selected only
by their photometric parallaxes. No significant differences were found,
but there remains the likelihood that some low proper motion objects
remain undiscovered. The second issue was binarity. Two LFs can be
constructed; one with the luminosity of ``systems'', the second with
the luminosity of the components of those systems. Binarity has been
subjected to extensive investigation in the PMSU survey, but less
attention has been given to the brighter Hipparcos sample! Reid \etal\
coped with this by assuming that the known binary population 
in the brighter stars is representative of the whole sample and simply
assigning these stars double weight in the analysis to deal with an
assumed factor of two incompleteness in the census. Despite this
additional assumption, knowledge of the binary fraction and companion
properties is a key advantage of any local population survey.

\begin{figure}
\begin{center}
\includegraphics[width=10cm]{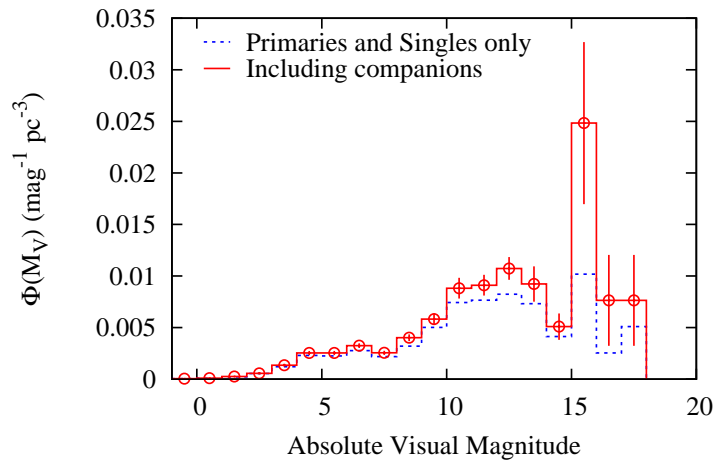}
\end{center}
\caption{The luminosity function of nearby stars from Reid \etal\
  (2002). The contribution of systems and all the components of systems
are shown separately.}
\label{nearbylf}
\end{figure}

The nearby star LF is shown in Fig.~\ref{nearbylf}, where the
``system'' and total LF are shown separately. Points to note are the
significant peak in the LF at $M_V \simeq 12$ (spectral type M3,
$\simeq 0.25\msun$). There is also a hint of the ``Wielen dip'' (Wielen 1974)
at $M_V \simeq 7.5$, which is now known to be caused by a kink in the
mass-magnitude relation caused by the details of changing opacities
in the outer envelope (Kroupa \& Tout 1997). 
Finally, note the large Poissonian error
bars that increase for $M_V>10$ (due to the small volume sampled at
these magnitudes) and that systematic uncertainties due
to unresolved or undiscovered binaries are also likely to become
important at these magnitudes.

Reid \etal\ convert this LF into a PDMF using two different
mass-magnitude relations (Kroupa \etal\ 1993; Delfosse \etal\ 2000), with
little difference in the results. It is worth noting that they used the
absolute
$V$ band magnitude in a population that must have a wide spread
of metallicities. From section~\ref{massmag} we understand that a near
infrared mass-magnitude 
relationship would be less metallicity sensitive and shows less
scatter, but if the calibrating binary population has a similar
metallicity distribution to the field population then the results should be
reasonably robust. 

To get from this observed PDMF to an IMF requires
corrections for (i) stellar evolution, (ii) the Galactic density distribution
and (iii) the mix of stellar populations that are included in the sample.
Reid \etal\ assume that the creation function $C(\log m, t)$ is
time-independent in which case the correction factor at any mass is
$\tau_{\rm G}/\tau_{\rm MS}$ (if $\tau_{\rm MS}<\tau_{\rm G}$). 
The second point arises because lower
mass stars are (on average) older, have a larger velocity
dispersion and hence a large scale height with respect to the Galactic
plane. For a density of the form $\rho = \rho_0 \exp(-z/z_0)$, then the
surface density of the disk is $\propto \rho_0 z_0$. Reid \etal\ model
this with a steep rise in scale height from $z_0=100$\,pc for $M_V<3$,
to $z_0=250$\,pc for $M_V>4$. Hence this is a very significant
correction and potentially there may yet be some mass dependence among
the lower mass stars. Finally, Reid \etal\ make a minor 10 per cent downward
correction for $M_V>4$, assuming that this fraction of the census
consists of very old, extended thick disk stars that make no
contribution to the IMF of the Galactic disk. 

The Reid \etal\ (2002) IMF of the Galactic disk is reasonably well
modelled between $0.1<m/\msun <3$ with a pair of power laws of the form
$\phi(m) \propto m^{-\alpha}$, with $\alpha \simeq 2.8$ for
$m>1.1\msun$ and $\alpha \simeq 1.3$ below this. The IMF shows no
significant evidence for a peak in the considered mass range (see
Fig.~\ref{bochanskimf}) and is less well represented by a log-normal
parameterisation (equation~\ref{lognorm}).

\subsection{Photometric field star surveys}

\subsubsection{Wide field and pencil beam surveys}

The alternative approach is to select a sample of low-mass stars from a
photometric survey; estimate their distances from a photometric
parallax; and then construct a LF that must be corrected for issues
such as binarity, Galactic structure and Malmquist bias (see below). 
These surveys can arbitrarily be divided between those that cover
relatively large areas on the sky, but reach comparatively shallow
apparent magnitude limits (e.g. Tinney \etal\ 1993; Covey \etal\ 2008;
Bochanski \etal\ 2010), and narrow, ``pencil-beam'' surveys going to
great depths (e.g. Gould \etal\ 1997; Martini \& Osmer 1998; Zheng \etal\
2001; Schultheis \etal\ 2006).
  
The earlier studies were focused on whether the $\alpha \simeq
2.35$ ``Salpeter law'' extended to very low masses, because this would
clearly have had a bearing on the question of baryonic dark matter. In
the event, these studies
appear to show that the slope of the IMF must fall to $\alpha
\simeq 1.0\pm 0.5$ somewhere between 0.5\msun\ and 1.0\msun, 
in agreement with the nearby star IMF, and so low-mass stars are not a
significant contributor to dark matter.

\subsubsection{A case study}

The recent work by Bochanski \etal\ (2010) makes an excellent case
study. This consisted of magnitude- and colour-selected samples from
8400 square degrees of the SDSS. After cleaning out bad photometry and
galaxies, the sample consists of 20 million late K- and M-type dwarfs.
As the depth limit of $r^\prime <22$ is
sufficient to probe well above the Galactic plane, the survey is both
deep enough and wide enough to encompass problems that affect both
general types of photometric field star survey.

Bochanski \etal\ use several colour-magnitude relationships in order to
estimate distances to their sample stars. These relationships were
calibrated in the SDSS photometric bands using stars with very precise
($\sigma_\pi/\pi <0.1$) trigonometric parallaxes. The typical scatter is
0.4\,mag, which is important when considering Malmquist bias (see below).
For each half-magnitude interval, the stellar density is calculated as a
function of position with respect to the Sun. These densities were
matched to a 2-D Galactic model; the free parameters consisted of
vertical and radial scale heights for both thick and thin disk
components, $f$, the fraction of the sample belonging to the thick
disk population, and a normalisation equal to the local density at the
solar position. The values of local density are the raw measurement of
the LF that then needs correcting for systematic effects.

The colour-magnitude relationships have a poorly calibrated metallicity
dependence and this combined with the expected metallicity gradient as
a function of height above the Galactic plane lead to systematic
differences in the LF -- the scale heights become smaller and the local
densities a little larger. For plausible metallicity gradients this
effect is small.

Extinction is modelled using a Galactic dust model. The dominant effect
is the reddening of stars rather than attenuation, which causes
absolute magnitudes to be underestimated. Most of the stars in the
sample probably have very small extinction ($A_r < 0.4$) and so even if
extinction is ignored entirely it only increases the LF by $\sim 10$
per cent among the brighter stars.
 
\subsubsection{Malmquist Bias}

The luminosity function of a magnitude-limited sample, or one where
distances have been estimated from photometric/spectroscopic parallax,
needs to be corrected for Malmquist bias (Malmquist 1936). 
This arises because distant
stars with brighter absolute magnitudes (at a given colour or spectral
type) scatter into the survey volume either because of the intrinsic
dispersion in the colour-magnitude relation or because of observational
uncertainties. A classical expression for the Malmquist bias, assuming
a Gaussian dispersion $\sigma$ in the colour-magnitude relation, is
\begin{equation}
\overline{M} = M_{\rm true} - \frac{\sigma^2}{N({\rm mag})}\frac{dN({\rm
      mag})}{d{\rm mag}} \, ,
\end{equation}
where $\overline{M}$ is the mean absolute magnitude estimate, $M_{\rm true}$
is the true mean absolute magnitude, and $N({\rm mag})$ is the number of stars
per unit {\it apparent} magnitude. For a uniform spatial distribution
of stars this
leads to the well known result that $\overline{M} = M_{\rm true} -
1.38\, \sigma^2$ (e.g. Butkevich \etal\ 2005). In turn, this correction
affects the observed LF because the sampled volume is effectively
increased and also, if $\Phi(M)$ has a steep slope, the LF is being
sampled at a value $M<M_{\rm true}$.

In non-uniform samples the Malmquist bias is more complex. 
Bochanski \etal\ (2010) deal with it using Monte Carlo simulations to
inject the appropriate scatter into the absolute magnitudes and colours
and then recalculating the LFs. The result is a significant (10--40 per
cent), luminosity-dependent {\it
  decrease} in the LF (and hence MF). It is worth noting that if a
survey is sensitive enough to reach out to, and beyond, the edge of a spatial distribution -- for
example a very deep pencil beam survey out of the Galactic plane -- then the
Malmquist bias may be negligible (e.g. Gould \etal\ 1997).

\subsubsection{Binarity}

Binarity is another thorny issue in photometric surveys because usually the
sample stars are so distant that no sensible attempt to resolve binary
components is possible, and no systematic survey for RV variations will
have been carried out. This is certainly the case for the SDSS sample
studies by Bochanski \etal\ (2010). Unresolved binarity causes two main
problems in LF and MF determinations. First, the actual number of stars
present in the survey is greater than first thought and the masses of 
those stars will differ from a simple estimate based on the system
luminosity. Second, in a magnitude-limited survey, the observed LF and MF
are enhanced because binaries are brighter than single stars of the
same colour and can be seen to greater distances. A simple example
would be a survey containing only equal-mass binaries. There would be
twice as many stars as there are measured stars in the survey, but they
will gathered from $2\sqrt{2}$ times the volume that is estimated from
the colour-magnitude relationship for single stars. 

Bochanski \etal\ chose to construct two separate LFs, one corresponding
to whole systems and the other to components of those systems
(including single stars). To calculate the component LF, a 
fraction of binary systems is added to
the original stellar catalogue. These have secondary component
luminosities (and corresponding colours) drawn randomly from the system
LF, but constrained to be fainter than the primary. The flux and
combined colour of each pair is then calculated. This new stellar
catalogue is passed through the analysis pipeline again to determine a
new system LF and this is compared with the initial system LF. The
initial system LF is then ``tweaked'' and the process iterated until
the artificially generated system LF matches that which was observed.

Dealing with binarity will always be a weakness of
unresolved field star surveys because the binary properties of the
lowest mass stars are rather poorly known. The binary frequency
is likely to decrease from $\sim 50$ per cent at a solar mass to about
10--20 per cent at the lowest stellar masses (see Duquennoy \& Mayor 1991;
Fischer \& Marcy 1992; Allen 2007 and Artigau's contribution in this
volume).  Bochanski \etal\ (2010) tried a variety of binary
prescriptions, with both fixed binary frequencies of 30--50 per cent
and also a frequency that declined with decreasing luminosity between
these two limits.  These differing assumptions play a significant role
in estimating the LF of the component stars, especially at the lowest
luminosities (and masses, $m<0.3\msun$), because many stars will have
unresolved companions at these masses (see Fig.~21 in Bochanski \etal\
2010).

\subsubsection{Results and parameterisations}

\begin{figure}
\begin{center}
\includegraphics[width=10cm]{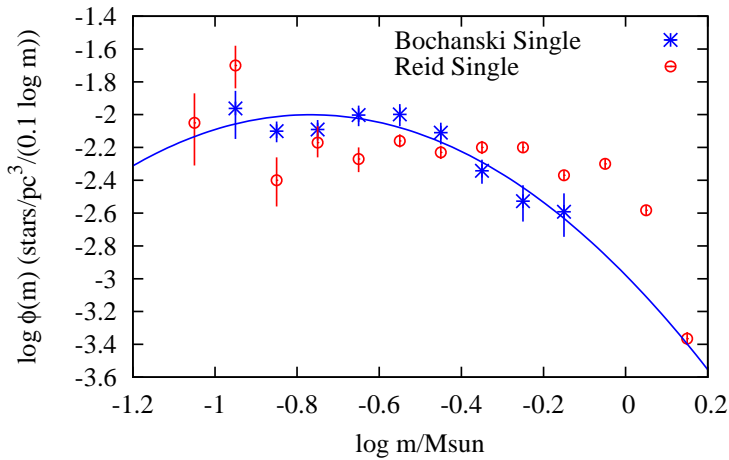}
\end{center}
\caption{The initial mass function for stars (single stars plus components of
  binary systems) based on a local sample (from Reid \etal\
  2002) and derived from a deep photometric survey (Bochanski \etal\
  2010). The IMFs have been corrected for stellar evolution (for
  $m>1\msun$) and are based on the local density of stars. The solid
  line is a log-normal fit to the Bochanski \etal\ IMF, with $m_c =
  0.18\,\msun$ and $\sigma=0.34$\,dex (see equation~\ref{lognorm}).
}
\label{bochanskimf}
\end{figure}

Bochanski \etal\ (2010) convert their LFs to MFs using the J-band LF
combined with the mass-magnitude relationship from Delfosse \etal\
(2000). The results for single stars plus the components of binary
systems are reasonably comparable with the IMF from local
field stars (Reid \& Gizis 1997; Reid \etal\ 2002) and are best fitted
with a log-normal mass function (see equation~\ref{lognorm}) with
$m_c=0.18\msun$ and $\sigma=0.34$\,dex, though a broken power law with
$\alpha =2.38$ for $m>0.32\msun$ and $\alpha=0.35$ for $0.1<m/\msun
<0.32$ is almost as good (see Fig.~\ref{bochanskimf}). 
The system IMF is fit with $m_c=0.25\msun$ and $\sigma=0.28$\,dex. In fact, the
evidence for any peak in the IMF is very weak on the basis of this
sample alone; the peak of the fitted function appears to coincide with a slight
(but insignifcant) dip in the IMF found from both local and extended samples.
Note that whilst the statistical uncertainties
in the Bochanski IMF are extremely small, the contribution of the systematic
uncertainties (extinction, metallicity dependence, binarity, the
Galactic model, Malmquist bias) means that the final estimated uncertainties
are as large as the Poissonian uncertainties in the IMF
determined from small local volumes, and actually larger for the higher
masses.

Compared to the IMF from nearby stars there is an excess of stars in
the Bochanski IMF for for $0.15 < m/\msun<0.4$ and a deficit for
$m>0.6\msun$. A suggested explanation for the deficit is neglecting to
include lower mass binary companions to stars with $m>0.9\msun$ that
are not included in the survey (e.g. Parravano \etal\ 2011). Certainly,
the width of the fitted log-normal functions are considerably narrower
than the $\sigma=0.55$\,dex suggested by Chabrier (2005) for the field IMF,
although the peak mass is similar.

\subsection{Summary of the Field Star IMF}
\begin{figure}
\begin{center}
\includegraphics[width=10cm]{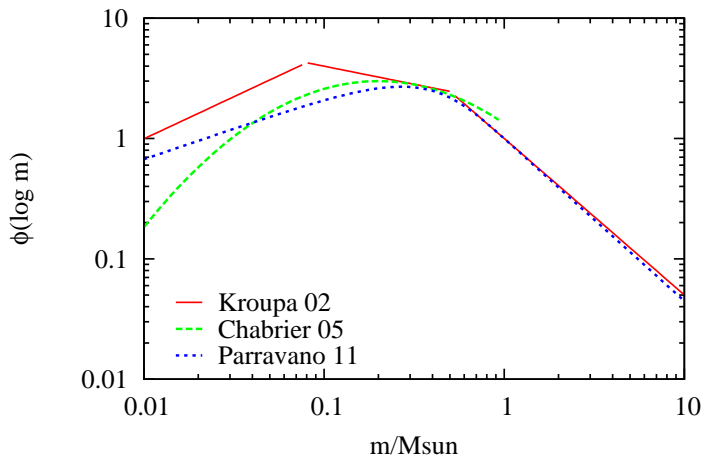}
\end{center}
\caption{A comparison of the different functional forms that have been
proposed to describe the IMF of the Galactic disk. The functions have
been normalised to be similar at 0.4\msun\ where the observational data
are most constraining. Above 1\msun\ the IMF is Salpeter-like, but
significant differences between the different parameterisations arise
for $m<0.2\msun$.}
\label{paramcompare}
\end{figure}

Assuming that the PDMF is
almost equal to the IMF readily allows its determination for stars with
$0.1<m/\msun<0.8$  in the general field
population. This can be done using
volume-limited local surveys or larger, magnitude-limited photometric
surveys. The uncertainties in local surveys are dominated by small
number statistics at low masses, whilst large scale photometric surveys are
dominated by systematic uncertainties that include the metallicity
dependence of photometric parallax relations, extinction, Malmquist bias and
corrections for binarity.

The IMF from both types of survey is converging to a uniform
result. The IMF can be represented in terms of more than one power law
(Kroupa 2002; Parravano \etal\ 2011) or a log-normal function with
characteristic mass of $0.15$--$0.25\msun$ (Chabrier 2005). The key
point that seems almost universally agreed is that the IMF flattens
significantly from a Salpeter-like slope at masses of 0.5--1\msun.
That there is little agreement on which parameterisations is
best is hardly surprising. When plotted such that they are normalised
to be equal at $\simeq 0.4\msun$, there is little difference
between them in the range $0.2<m/\msun<0.8$ where the data are most
secure (see Fig.~\ref{paramcompare}). There are however clear
differences at lower masses and especially in the brown dwarf regime
and it is to the IMF of BDs that we turn next.

\section{The Mass Function for Brown Dwarfs}

\label{lec3}

The IMF of low-mass stars is estimated by construcing a LF and then
converting this to a MF using mass-magnitude (or mass-luminosity)
relationships that are (i) quite well calibrated and (ii) not very age
dependent. This section moves to lower masses ($m<0.1 \msun$), where
neither of these properties can be assumed. The age-dependent
mass-luminosity relationship demands a different, inductive
approach. Despite these difficulties, enormous efforts have been
expended on this task because there is hope that the form of the IMF at
low-masses will not only solve the problem of which parameterisation
best represents it (see Fig.~\ref{paramcompare}), but more importantly
will actually yield physical insight into which mechanisms control the
form of the IMF, determine it's characteristic mass and what {\it
might} be responsible for IMF variations (e.g. Padoan \& Nordlund 2004;
Hennebelle \& Chabrier 2008). In this section I explain the
methodologies adopted and discuss the results.

\subsection{The mass-luminosity relation for brown dwarfs}

BDs are very low-luminosity objects with no nuclear source of
heat (bar a brief deuterium-burning phase if $m>0.015\msun$). As
field objects they were first identified as companions to low-mass
stars (Nakajima \etal\ 1995), and even in the near-IR are $>7$ mag
fainter than solar-type stars. The brightness contrast is less severe
for young brown dwarfs and another fruitful place to find them has
proved to be young clusters like the Pleiades and the Orion Nebula
cluster (see section~\ref{lec4}; Rebolo \etal\ 1995; Lucas \& Roche 2000).
Searching for isolated objects in the field with $m<0.1\msun$ poses
significant problems that have only recently been addressed.

The mass-luminosity relation of BDs was discussed in
section~\ref{bdml}, but it is worth re-iterating the main points.
First, the mass-luminosity relation of BDs is age-dependent. That
is, even were the evolutionary models absolutely accurate, then the age
of a BD is required before one can estimate a mass from its luminosity.
Second, the accuracy of the models cannot be assumed. There is already
a lot of evidence that the theoretical evolutionary models and the
model atmospheres do not correctly predict luminosities of BDs of a 
given mass and age, and they do not correctly reproduce the positions of
BDs in the Hertzsprung-Russell diagram. To quote King \etal\ (2010) from
their work on the BD binary $\epsilon$ Indi Bab -- ``Given the
preliminary dynamical system mass it therefore appears that with
current theoretical models and spectroscopically derived
effective temperatures, one cannot obtain reliable mass estimates for
T-dwarfs such as these, even when precise luminosity constraints are
available.'' With that depressing assessment, we turn to methods of
deducing the IMF for BDs using these very models!

\subsection{Simulating field brown dwarf luminosity functions}

The age-dependent mass-luminosity relationship means that determining
the IMF from an LF is inextricably linked to the age distribution of the
population in question and hence to the star (BD) formation
history. Several authors have suggested an inductive approach that
predicts what the LF (or the distribution of effective
temperatures/spectral types) of a low-mass field population will be given
assumed functional forms for the IMF and star forming history.
The advantages of this approach (in contrast to work on young clusters
discussed in section~\ref{lec4}) are that the field BDs will be close and
more amenable to determining their binary characteristics, and it is
hoped that, despite the limitations discussed above, the
evolutionary models may be more accurate for older BDs.
Burgasser (2004) gives a very clear description of the
general approach; 
the conclusions it reaches and the limitations of such work are
similar to work by Reid \etal\ (1999), Allen \etal\ (2005), Deacon \& Hambly
(2006) and Burgasser (2007).

In Burgasser (2004) IMFs of the form $\phi(m) \propto m^{-\alpha}$ are
considered along with the log-normal form suggested by Chabrier
(2003). These IMFs cover objects with $m<0.1\msun$; the normalisation
was set to match local field stars (both single and in multiple
systems) with $0.09<m/\msun<0.1$, noting that this scaling factor is
still uncertain by 30 per cent. Considered star forming histories
includine a uniform distribution, an exponentially decaying function
and various ``burst'' models. It is assumed, as usual, that the IMF is
time-independent. Monte Carlo simulations, select objects at random
from the age distribution and from the IMF with masses between
0.1\msun\ and a number of possible low-mass cut-offs. The present day
properties of the objects were derived using the models of Burrows
\etal\ (1997) and Baraffe \etal\ (2003) and these were binned to form
luminosity and $T_{\rm eff}$ distributions.

The main limitations of this approach are the absolute reliance on
evolutionary and atmospheric models to give the correct properties for
BDs of a given mass and age. Presumably the local field population of
BDs spans a range of metallicities, but the metallicities of field L-
and T-dwarfs are not well known and the evolutionary models available
all assumed solar metallicity.

The key results of these simulations, which are 
confirmed by the later works, are (see Figs.~9--12 in
Burgasser 2004):
\begin{enumerate}

\item The LFs and $T_{\rm eff}$ distributions are morphologically
  similar, because most BDs with ages of 1--10\,Gyr have radii close to
  1\,$R_{\rm Jup}$. 
  There is a peak among late M-dwarfs caused by
  long-lived 0.08--0.1\msun\  stars; a trough through the L-dwarf
  regime ($1400<T_{\rm eff}<2300$\,K) because the evolutionary models
  show that BDs with $m<0.075\msun$\ cool rapidly at these
  temperatures; there is a ``pile-up'' of cooling T-dwarfs, 
  rising to a peak in late T-dwarfs because the cooling time of such
  objects reaches 10\,Gyr  (see Fig.~\ref{burglf}).

\item The LFs and $T_{\rm eff}$ distributions become
  increasingly sensitive to variations in the adopted IMF at
  lower luminosities and cooler temperatures. There are factors of a
  few variations in the late T-dwarf LF for
  a range of plausible IMFs (see Fig.~\ref{burglf}). If $\alpha=1$,
  there should be $\sim 60$ T6--T8 dwarfs within 10\,pc of the sun,
  but only $\sim 25$ if $\alpha=0$.

\item There is little dependence on adopted evolutionary model for
  spectral types cooler than late-L. This must be tempered by the fact that the
  evolutionary models do have some common assumptions and presumably
  weaknesses. There is some model-dependence (at the level of factors
  of two) for late M and early L-dwarfs.

\item There is only a modest dependence on the birthrate model.
Any dependence is confined to L-dwarfs. In particular, the space
density of $\geq$T6 dwarfs is insensitive to the star forming history of the
Galactic disk.

\item Only the space density of the very coolest T-dwarfs (T8--Y0?) is
  sensitive to any cut-off or minimum mass in the BD IMF if that
  cut-off is $<0.015\msun$.

\end{enumerate}

In summary, the space density of late T-dwarfs is most sensitive to the
detailed shape of the BD IMF, but is almost insensitive to the star
forming history. The present day density of L dwarfs may offer some
sensitivity to the BD birthrate history, but it will be difficult to
disentangle this from systematic differences of similar size between
the predictions of alternative evolutionary models.

A final complication is the influence of binarity on these
simulations. The binary properties of BDs are quite uncertain. Resolved
BD binaries among the nearby field population amount to about 15--30
per cent of the observed population (see Table~1 in Burgasser 2007), with
evidence of higher binarity among early L8--T3 dwarfs, which may be
consistent with an underlying (resolved) binary frequency of
$11^{+6}_{-3}$ per cent. Burgasser (2007)
has conducted a further series of simulations incorporating the binary
population, predicting what would be observed among volume-limited
or magnitude-limited surveys with no binary resolution. The effect is
small in volume-limited surveys, but if, as is more usual, a survey is
magnitude-limited then there can be significant corrections because of
the additional volume from which binaries can contribute.

\subsection{Measuring field brown dwarf luminosity functions}

Armed with predictions from simulations, it is merely(!) a case of
conducting a census of local BDs, grouping them into luminosity or
temperature bins and comparing them with the models.  The challenges
involve (i) finding cool BDs and defining a clean sample, (ii)
assessing completeness, (iii) assessing contamination, (iv) determining
distances to calculate a space density and (v) correcting these
densities for Malmquist bias and the presence of binary systems.  A
number of major surveys have been brought to bear on this topic. Key
literature includes Allen \etal\ (2005), Cruz \etal\ (2007), Metchev
\etal\ (2008), Reyl\'e \etal\ (2010) and Burningham \etal\ (2010), that
have used a variety of survey data and photometric colours to select
samples of cool BDs (see Table~\ref{bdsurvey}).

\begin{table}
{\small 
\begin{tabular}{l@{\hspace{2mm}}c@{\hspace{2mm}}c@{\hspace{2mm}}c@{\hspace{2mm}}c@{\hspace{2mm}}c}
\hline \\
Study &  Survey & Method& Sample & $\alpha$ & Mass range \\
\hline \\
Allen   & 2MASS&vol-lim(ML)& 180 ML &
 $+0.3\pm0.6$ & $0.04$--$0.10\msun$\\
\etal\ 2005              & 14800 deg$^2$ & mag-lim(T)& 18 T5-T8&
& \\
 & $J<16.5$ & $JHK+$ spec. &&&\\
\\
 Cruz   & 2MASS                   & vol-lim    & 45 L &
 $\leq 1.5$ & $\leq 0.075\msun$ \\
 \etal\ 2007            & 14800 deg$^2$  & $JHK+$spec.  &      &
	    &                    \\
& $J<16.5$ &&&& \\
\\
Metchev & SDSS/2MASS & mag-lim & 15 T0-T8 & $\leq 0$ & $\leq
0.075\msun$ \\
\etal\ 2008    & 2099 deg$^2$ & $izJHK+$spec. & & & \\
& $z<21$ &&&& \\
\\
Reyl\'e  & CFBDS  & mag-lim & $102$ L5-T9& $\leq 0$ &$\leq
0.075\msun$ \\
\etal\ 2010    & 444 deg$^2$ & $izJ+$spec. & & & \\
& $z<22.5$ &&&& \\
\\
Burningham & UKIDSS & mag-lim & 47 T6-T9& $-0.5\pm0.5$ &
0.02--0.04\msun \\
\etal\ 2010   &980 deg$^2$ & $zYJH+$spec. & & & \\    
& $J<18.8$ &&&& \\
\hline \\
\end{tabular}
}
\caption{A summary of surveys for the local field BD
  population. Columns list the sample source (see text), magnitude
  limit, size of sample, the deduced power-law index of the IMF
  (equation~\ref{powerlaw}) and the sampled mass range.}
\label{bdsurvey}
\end{table}

It can prove difficult to separate very low-mass
objects from the higher mass M-dwarfs that dominate at any
magnitude.
The optical spectrum of M-dwarfs is dominated by TiO bands, but these
become progressively weaker and disappear by spectral type L3. At the
same time a number of atomic lines (rubidium, caesium) become broad and
prominent, while the potassium 7665/7699\AA\ line becomes a broad trough in
the spectrum at L5. In the near infrared, water absorption appears in
M-dwarfs and grows in strength. At around 1400\,K, methane absorption
becomes prominent (defining the T-dwarf class) and significantly
alters the near infrared spectrum. The effects of these
spectral changes on commonly used photometric colours is well
illustrated 
in Fig.~4 of Chiu \etal\ (2006). L dwarfs are a little redder than
M-dwarfs in all colours and hence many candidates are found in 2MASS
and SDSS data. However, the separation in colour from M-dwarfs is
insufficient to prevent photometrically selected samples being heavily
contaminated by the much more common M-dwarfs, 
because both intrinsic colour dispersion and
photometric uncertainties scatter them into the selection region.
The situation is more favourable for T-dwarfs, because while being
redder than L-dwarfs in $i-z$ and $z-J$, they are significantly bluer
at $J-H$ and $H-K$. Hence whilst late L- and T-dwarf candidates can be
selected in (for instance) $i-z$ vs $z-J$ diagrams (e.g. Metchev \etal\
2008; Reyl\'e \etal\ 2010), the addition of $H$ or $K$ data appears to
offer much more discrimination for T-dwarfs (Burningham \etal\ 2010).

Attempts to determine spectral types from photometry are almost
impossible. The intrinsic dispersion of colours at a given spectral
type can reach 0.5 mag. The reasons for this are poorly understood but
probably include the presence of unresolved binary systems (especially
around the L/T transition), differences in metallicity, patchy
clouds/condensates and a range of surface gravities because objects at
a given $T_{\rm eff}$ have a range of ages and masses (e.g. see Knapp
\etal\ 2004; Liu \etal\ 2006; Burgasser 2007; Marley \etal\ 2010). This
means that detailed comparisons with simulations demand spectroscopic
spectral type determination.

Both completeness and contamination are issues to be dealt with in
these studies. Contamination arises because numerous unwanted objects are
present or are scattered into the photometric selection box. This can
be dealt with either by detailed simulations (Reyl\'e \etal\ 2010) or
spectroscopic follow-up to eliminate contaminants.  The simulation
technique requires an
accurate knowledge of the (possibly non-Gaussian) tail of the
photometric uncertainty distribution.  Completeness corrections are
required to deal with the sensitivity of the observations, but also to
account for any possibility that photometric dispersion can also
scatter genuine candidates {\it out} of the selection box. Different
approaches to this include making a correction based on the recovery of
previously known objects (Cruz \etal\ 2007), or simply ignoring
spectral type ranges where the photometric selection box may exclude
genuine BDs (Burningham \etal\ 2010).

After settling on a census then distances are estimated.  Parallaxes
are rarely available so relationships between colour, or preferably
spectral type, and absolute magnitude are used, calibrated using some
objects with known (and precise -- see section~\ref{lutzkelker})
parallax (e.g. Cruz \etal\ 2003 for late M- and early L-dwarfs or Vrba
\etal\ 2004 for L- and T-dwarfs). Like the spectral type versus colour
relationships, there is a dispersion, possibly caused by gravity and
metallicity differences. There is also a pronounced kink or plateau for
early T dwarfs, which may partly be explained by binarity (Liu \etal\
2006). Different authors use different forms of these relationships
that can lead to factors of two systematic differences in their
estimates of space densities. These space densities then have to be
corrected for the significant Malmquist bias due to dispersion in the
spectroscopic parallax.

Studies also differ in how they treat binarity. The effect of binarity
on volume-limited LFs is small because the binary frequency of BDs is
low. Nevertheless, a significant correction to a LF derived from a
magnitude-limited survey is still required because the volume sampled
by binary systems can be up to $2\sqrt{2}$ larger (for a mass ratio of
1) than for single BDs.  Corrections have been calculated using
analytical approximations (e.g. see Burgasser 2004; Burningham \etal\
2010), by Monte Carlo simulation (Metchev \etal\ 2008), {\it reducing}
the LFs by up to a factor of two, or have been neglected (Reyl\'e
\etal\ 2010).
 
\subsection{The IMF of field brown dwarfs}

\begin{figure}
\begin{center}
\includegraphics[width=11cm]{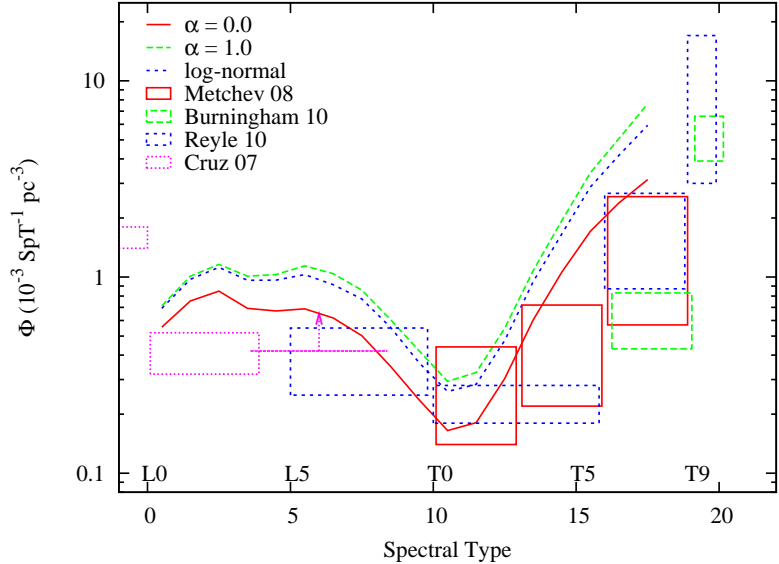}
\end{center}
\caption{The luminosity function of field BDs as a function of their
  spectral type. Results are shown for a number of recent surveys (see
  text) Also shown are model loci taken from Burgasser (2007). These
  predict the observed LF based on a set of evolutionary models and
  atmospheres together with an assumed value for the IMF ($dN/dm 
  \propto m^{-\alpha}$) and a uniform formation rate. Also shown is a
  locus derived using a log-normal IMF. Overall it appears that values
  of $\alpha<0$ are favoured, though a log-normal IMF with a smaller
  dispersion parameter may also be possible (see text).
}
\label{burglf}
\end{figure}

The final LFs are compared with simulated LFs predicted
according to different assumptions about the IMF and birthrate history.
Before doing so it is worth pausing to recall the systematic
uncertainties present. The most important are: (i) The veracity or
otherwise of the
evolutionary models and their predictions of luminosity and $T_{\rm eff}$ as a
function of mass and age. (ii) A relationship between $T_{\rm eff}$
and spectral type is needed to make an observational comparison
(e.g. Golimowski \etal\ 2004). Systematic uncertainties in this will
redistribute objects between spectral types. (iii) Systematic
uncertainties in spectroscopic parallaxes could change space densities
by factors of 1.5--2. (iv) The model normalisation at 0.1\msun\ is
uncertain by about 30 per cent. (v) Corrections for binarity are
uncertain at a similar level.

A comparison between measured and predicted LFs is shown in
Fig.~\ref{burglf}. The model loci are from Burgasser (2007) and assume
a uniform star forming rate, a local field density 
normalisation of 0.0037 pc$^{-3}$ for
stars with $0.09<m/M_{\odot}<0.1$ (Reid \etal\ 1999), the evolutionary
models of Burrows \etal\ (2001) and Baraffe \etal\ (2003) and the
``COND'' models of Allard \etal\ (2001). A variety of IMFs of the form
$dN/dm \propto m^{-\alpha}$ were explored by Burgasser, and also a
log-normal IMF with parameters taken from Chabrier (2002).
Subject to all the caveats listed above, it seems that the current
observational data favours a value of $\alpha \leq 0$. The log-normal
model of Chabrier (2002) does not describe the observations well but
the parameters of this distribution, $m_c = 0.1\,M_{\odot}$ and $\sigma
= 0.627$\,dex (see equation~\ref{lognorm}) were updated to
$m_c=0.2\,M_{\odot}$ and $\sigma=0.55$\,dex by Chabrier (2005) and may
describe the observations significantly better.

\section{The initial mass function from young clusters}

\label{lec4}

\subsection{Advantages and disadvantages}

\begin{figure}
\begin{center}
\includegraphics[width=10cm]{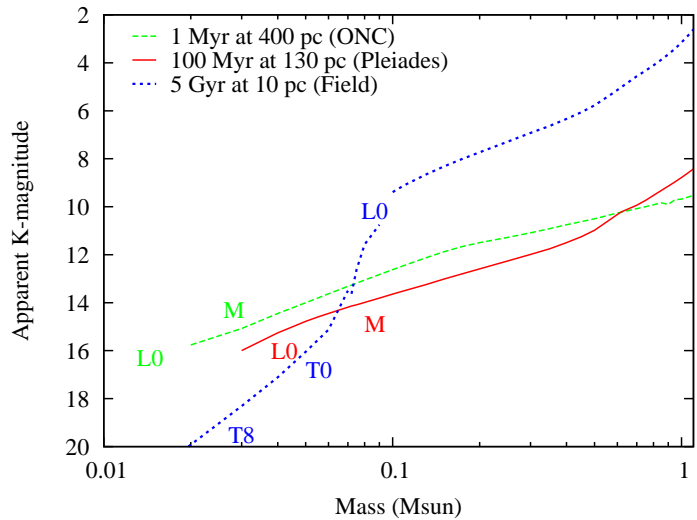}
\end{center}
\caption{The apparent $K$ magnitude versus mass for low-mass objects
  for ages/distances representative of young clusters and the field.
  The loci are taken from the models of Baraffe \etal\ (2002), with the
  models for cooler field brown dwarfs taken from Chabrier \etal\
  (2000, for $T_{\rm eff}<2700$\,K, the DUSTY models) or Baraffe \etal
  (2003, for $T_{\rm eff}<1700$\,K, the COND models). Even nearby field
  brown dwarfs are fainter than younger objects of similar mass in the Pleiades and
  the Orion Nebula cluster (ONC).} 
\label{kvsm}
\end{figure}

An alternative to finding low mass stars and BDs in the field is to
look for them in star clusters. 
The primary scientific advantage of determining the IMF from clusters
is that whilst the local field IMF is some sort of average over all
star forming environments, clusters offer the opportunity to probe
differing star forming conditions (e.g. Jeans mass, density, radiation
environment). Open clusters contain approximately coeval stars with
similar initial composition. In principle the ages of clusters are
known, either from the main sequence turn-off at high masses or from
other techniques, so the difficulties of a time-dependent
mass-luminosity relationship are resolved. Distances can also be
determined by main sequence fitting or other means, 
and even though clusters are
usually more distant than field BDs identified in wide
photometric surveys, their youth means that low mass objects in
clusters are hotter and intrinsically more luminous. The net result is
that despite their distance, BDs in clusters 
can be brighter than field objects of similar mass (see Fig.~\ref{kvsm}).
 
In the debit column, very young clusters ($<10$ Myr) are afflicted by
considerable and variable extinction that makes luminosities hard to
estimate. Very young objects are often surrounded by circumstellar
accretion disks which alter their spectral energy
distributions making an estimate of the underlying luminosity (and
hence mass) problematic (Hillenbrand 1997; Da Rio \etal\ 2010).  The
masses must be estimated using theoretical models but these have not
been well-tested at at younger ages and are likely to be affected by
significant uncertainties (e.g. Baraffe \etal\ 2002, 2009).
At very young ages a small age spread could make a significant
difference to the mass-luminosity relation. The reality or not of
significant age spreads is currently disputed (e.g. Huff \& Stahler
2006; Jeffries \etal\ 2011).

In older clusters coevality must be a good approximation, there is also
little variable extinction and no accretion. But the measured MF may no
longer be the IMF.  Mass segregation effects that are either primordial
(therefore affecting young clusters too) or which develop dynamically
as clusters gets older can alter the MF significantly. Care must be
taken that a census of cluster members is not biased by any radial
dependence of the MF.  Some fraction of the lower mass objects may be
preferentially evaporated from the cluster resulting in the PDMF
underestimating their contribution (Allison \etal\ 2009; de Marchi
\etal\ 2010). As most cluster MFs make no attempt to resolve binary
systems, a more subtle effect could be any change in binary frequency
with age or environment (e.g. Kroupa \& Bouvier 2003).  In these
circumstances there may be little alternative but to appeal to
simulations in an effort to predict what the IMF was.

Whichever type of cluster is being observed, to perform an accurate
census there are two issues that must always be considered;
contamination and completeness.  Contamination of samples, whether they
be selected on the basis of colours, proper-motions or radial
velocities will usually be a problem.  Ideally one would like to work
with samples that have been cleaned of significant contamination
(e.g. background/foreground field stars, background galaxies and
quasars) by applying multiple membership criteria or through
independent membership criteria. The same selection criteria may also
lead to incompleteness in the sample if they are too stringent. In
addition, photometric surveys are usually incomplete at some level due
to magnitude or colour limits, variable extinctions and obscuration by
brighter stars or nebulosity.

\subsection{Cluster Membership}

\subsubsection{Photometry}

The first (and sometimes only) line of attack in cluster MF determinations
 is to examine colour-magnitude or colour-colour diagrams to
 select cluster members and exclude non-members.
In the absence of age spreads, coeval cluster members of uniform
 chemical composition are expected to follow well-defined loci in such
 diagrams.
  Sequential selection
 in a number of diagrams can be made but there is almost always some
 contamination that remains (e.g. Lodieu \etal\ 2009; B\'{e}jar \etal\
 2011). The selection boundaries must be broad
 enough to include all possible cluster members (including binary or
 possibly even higher multiple systems) or the selection criteria must
 be well enough understood that the mass-dependence of any
 incompleteness can be accounted for at a later stage. There is
 a tension between ensuring completeness and minimising contamination.
 The amount of contamination depends on the depth of the survey, the
 photometric bands being used and the Galactic longitude and latitude
 of the cluster. A variety of techniques can be used to estimate
 contamination based solely on the photometric information or by using
 semi-empirical models for the foreground and background Galactic
 populations (e.g. see Jeffries \etal\ 2004; Oliveira \etal\ 2009).

The selection region for cluster members is often defined with
reference to theoretical isochrones if the cluster age is
well-known. However, this approach is susceptible to uncertainties in
the theoretical models and atmospheres, a particular hazard for very
low-mass stars and brown dwarfs. Guidance is often taken from the known
location of cluster members that have been confirmed using alternative
techniques (RVs, proper motions, Li abundances etc.). 

In general, photometric selection of cluster members works better in
older clusters where there is a negligible spread in luminosity caused
by any plausible age spread, no accreting objects with peculiar
luminosities, and often a uniform extinction. For selecting very
low-mass stars and brown dwarfs it is almost mandatory to have data at
infrared wavelengths, although selection in $I, I-Z$ or $I, R-I$
diagrams can be effective.

\subsubsection{Kinematic selection}

Cluster members are often assumed to share a common space velocity.
Measurements of the proper motions (PMs) and RVs of clean samples of
members in older well-populated clusters suggest velocity dispersions
of about 1\,km\,s$^{-1}$ for most open clusters (e.g. Geller \etal\
2010; Jackson \&
Jeffries 2010; Bonatto \& Bica 2011).  If clusters are virialised then
equipartition would give a larger velocity dispersion to lower mass
stars.  Individual clusters show signs of both sub- and
super-virial motions, reflecting details of their formation or
dissolution processes (e.g. de Grijs \etal\ 2008; Proszkow \etal\
2009). Some formation models for very low-mass
stars and BDs predict much higher velocity dispersions for these
objects (Reipurth \& Clarke 2001; Umbreit \etal\ 2005). 
With these caveats in mind, kinematic observations provide a
useful, though rarely decisive indication of cluster membership.

The utility of PM measurements is a function of their precision,
the distance to the cluster, the peculiar tangential motion of the
cluster relative to the field population and the density of foreground
and background sources. Older, all-sky PM catalogues based on
photographic plates with long observation baselines have precisions of a few
mas/year, that are just capable of resolving cluster velocity
dispersions at $\sim 100$\,pc, but are limited to $V < 20$ which,
whilst useful in tracing the low-mass population struggles to
probe very low-mass stars and BDs in clusters any further
than this. Frequently, PMs are combined with photometric
constraints to provide a cleaner sample (e.g. Hambly
\etal\ 1999; Deacon \& Hambly 2004). The advent of new digital far-red and
infrared surveys such as 2MASS, SDSS and UKIDSS have improved the
situation considerably in nearby clusters 
(e.g. Adams \etal\ 2002;  Kraus \& Hillenbrand
2007; Lodieu \etal\ 2011). Nevertheless, unless the cluster population
dominates the field population in some limited area (e.g. NGC 3603,
Rochau \etal\ 2010), PMs can only be used
to reject cluster non-members in more distant clusters (e.g. Caballero
2010).

Different considerations apply to RV measurements. The
distance to the cluster is only relevant in that precise RV
measurements require at least intermediate resolution ($R\geq 5000$)
spectroscopy. The specificity of RV selection depends on the
cluster RV compared to that of the field
population. For clusters with distances of a few hundred
pc or more, greater membership 
discrimination can be achieved with RVs than with
PMs, and are facilitated by  wide-field fibre spectrographs
(e.g. Frinchaboy \& Majewski 2008), 
although challenging measurements in the far-red or near-infrared
are needed to reach BDs (e.g. Kenyon \etal\ 2005; Sacco \etal\ 2008) and
it is difficult to achieve high levels of completeness. Nevertheless,
the precision of such surveys often yield unexpected results such as
revealing that some clusters consist of multiple populations with
different ages and at different distances (e.g. Jeffries \etal\ 2006),
with obvious implications for IMF determinations!

\subsubsection{Youth indicators}

A variety of youth indicators are widely used to
identify members of young clusters -- which can be useful in
empirically defining where in CMDs the cluster
members lie.  All of these methods have their problems and are
unlikely to provide the sole means of completing a cluster census.

Magnetic activity is ubiquitous among young, low-mass stars and is
associated with their convective envelopes and rapid rotation. The
rapid rotation of young stars in clusters can be established either by
multi-epoch photometric monitoring, which determines rotation periods
from modulation by cool, magnetic starspots (e.g. Irwin \etal\ 2007;
Hartman \etal\ 2010) or by large-scale, high resolution spectroscopic
surveys that measure rotational broadening (e.g. Jackson \& Jeffries
2010). Unfortunately, only a fraction of young stars exhibit rotational
modulation in any survey and rotational broadening can be masked by
small rotation axis inclinations. An additional problem is that whilst
rapidly rotating field stars are rare among field G- and K-dwarfs, the
spin-down timescales become longer in M-dwarfs, and rapid-rotators
become common in field dwarfs cooler than M4 and in BDs (e.g.
Reiners \& Basri 2010; Reiners \etal\ 2012), so that incompleteness and
contamination of samples are problematic.

X-rays are emitted from the magnetically confined coronae of young,
low-mass stars (Guedel 2004).  X-ray imaging surveys can be used to
identify candidate cluster members and trace their spatial extent.
However, limited sensitivity means that these surveys rarely reach the
lowest masses, and BDs are probably weak X-ray sources except at vey
young ages (Grosso \etal\ 2007; Berger \etal\ 2010). IMFs derived
solely from X-ray selected samples would have uncertain levels of
incompleteness, especially in star forming regions where variable
extinction and accretion also serve to increase the scatter in observed
X-ray luminosity at a given mass (Flaccomio \etal\ 2010).

Chromospheric H$\alpha$ emission can also be used as a magnetic
activity indicator and is strong in the 
active K- and M-dwarfs of young clusters. The completeness levels of
any survey can be well-controlled and the observation of chromospheric
activity is a strong membership indicator in young K-stars but becomes
weaker in cooler stars because a large fraction of field M-dwarfs are
also magneticaly active. Emission persists until spectral types of at
least M8, but suffers a sharp decline at cooler temperatures
(e.g. Mohanty \& Basri 2003).

Indicators of circumstellar material or accretion are expected from
{\it a fraction} of very young low-mass stars. Examples include strong,
broad H$\alpha$ emission and infrared excesses radiated from
circumstellar dust. The former can be found either in wide-field
spectroscopic surveys, but can be investigated more economically
via photometric means (see Valdivielso \etal\ 2009 for an application
to very low mass stars and BDs).  Secure identification of infrared
excesses was revolutionised with 3--8\,$\mu$m observations by the Spitzer
satellite. Most stars have detectable circumstellar dust at ages of
1\,Myr, but this declines to a negligible fraction over the course of
10\,Myr (see Hern\'andez \etal\ 2008 and references therein).  Any
sample constructed from stars selected solely on the basis of an
infrared excess will be incomplete, although contamination levels will
be extremely low. An additional concern is that the level of
incompleteness may be mass-dependent. Although young BDs are also
found with discs and accretion, it is possible that their
timescales for disc retention are longer (e.g. Bouy \etal\ 2007).

\subsubsection{Spectral types and gravity-sensitive features}

In star forming regions it is important to establish spectral
types as a means of estimating extinction, separating photospheric
emission from accretion and hence establishing accurate luminosities.  Extinction
values can be used to select cluster members and it is often useful to
define an ``extinction-limited'' sample (i.e. a complete subset of the
full sample, with extinction less than some value) with which to
investigate the IMF; under the assumption that extinction and mass are
independent (e.g. Luhman 2007).

Low-mass PMS stars are larger than their main-sequence counterparts and
have lower surface gravity. At a given age, the difference is largest
for lower mass stars and leads to observable consequences in their
spectra. A prime example is the neutral alkali lines (K\,{\sc i},
Na\,{\sc i}) in the far-red spectrum.
These are weaker in cluster members than in main-sequence
dwarfs of similar colour or spectral type, but stronger than in giants
(see for example Luhman 2004; Lodieu \etal\ 2011).
As a method of identifying cluster members this is most effective in
M-dwarfs and can be done with relatively low-resolution spectra ($R>1000$). 

\subsubsection{Lithium depletion}

Lithium is burned in the cores of fully convective PMS stars as they
contract and heat up. Because convective mixing is effective, very
rapid and total Li depletion results except in higher mass stars
($m>0.6\msun$) where the genesis and expansion of a radiative core
occurs early enough to preserve some of the original Li. The rate of
PMS contraction is age-dependent so the amount of Li depletion depends
on both age and mass in a complex way (see Jeffries 2006), but
Li is undepleted for all low-mass stars ($m<0.25\msun$)
younger than 30\,Myr and for BDs of any age with $m<0.06\msun$. In
principle this is a powerful membership indicator in young clusters,
even for BDs because they can be separated from older Li-rich BDs by
their much higher temperatures (e.g. Rebolo \etal\ 1996; Zapatero-Osorio
\etal\ 2002; Kenyon \etal\ 2005). However, the technique requires
precise measurements of the Li\,{\sc i}\,6708\AA\ feature using
spectroscopy with $R>3000$.

\subsection{Results for Older Open Clusters}

\begin{figure}
\begin{center}
\includegraphics[width=10cm]{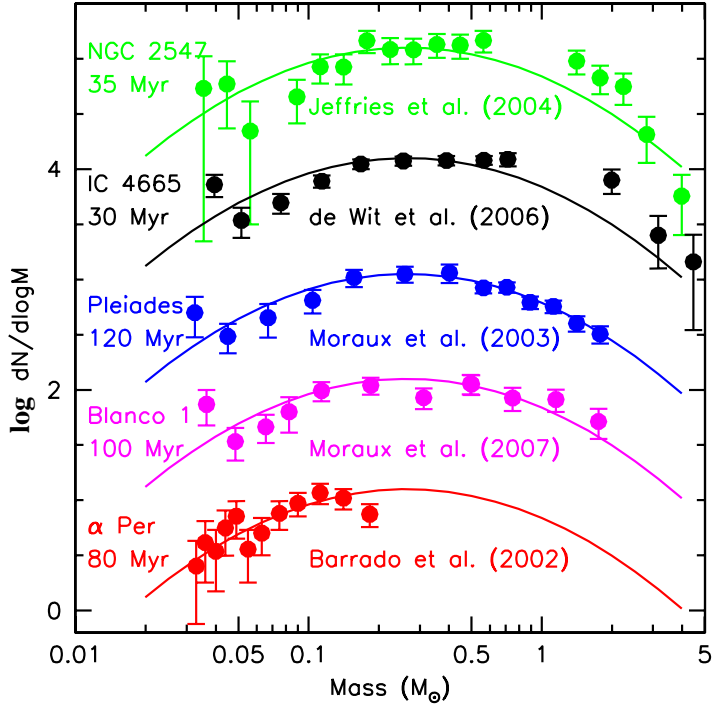}
\end{center}
\caption{The MFs for young clusters with ages 30--120\,Myr (offset for
  clarity). The solid
  line indicates the best-fitting log-normal description of the
  Pleiades MF ($m_c = 0.25\,\msun$ and $\sigma=0.52$\,dex). Figure
  constructed by Bouvier \& Moraux (private communication).
} 
\label{yocmf}
\end{figure}

Once a census has been taken then the masses of the stars and BDs must
be determined. The adopted process differs between open clusters with
age $>10$ Myr, where accretion, discs, variable extinction and age
spreads can be ignored, and the younger clusters and star
forming regions where they cannot. As a representative of the former, I
take the Pleiades as an example; a nearby cluster with a precisely
known age ($125\pm 10$\,Myr, Stauffer \etal\ 1998) and
distance ($135\pm 4$\,pc, An \etal\ 2007), 
with a well-established upper main sequence and mean proper motion.

The boundary between stars and BDs occurs at $I=17.8$, so relatively
deep far-red or infrared photometry is needed to locate its very low
mass members. Moraux \etal\ (2003) used an $IZ$ survey with ``optical''
CCD cameras in order to cover large areas (6.4 deg$^2$)
of the cluster. Selection in
the $I$, $I-Z$ CMD gave 40 candidates down to an assumed completeness
limit of $I\simeq 22$, $m\simeq 0.03\msun$. Contamination was estimated
to be about 30 per cent based on the LF and colours of nearby stars.
Masses were estimated directly from absolute $I$ magnitudes using
evolutionary models, with no attempt to correct for unresolved
binarity. The resulting MF was well represented either by a power law
with $\alpha = 0.60\pm0.11$ for $0.03\leq m/\msun \leq 0.45$ or a
log-normal form with $m_c=0.25$ and $\sigma=0.52$\,dex applicable for a wider
range of $0.03\leq m/\msun \leq 2$ (see Fig. \ref{yocmf}). Lodieu \etal\ (2007) used
``optical'' $IZ$, 2MASS infrared data and deep infrared photometry from the 
UKIDSS survey over 12 deg$^2$ to select 456 stars and BDs using
multiple CMDs and PMs. Membership probabilities are summed
to form an LF and this is converted to an MF using evolutionary
models. The results are in good agreement with Moraux \etal\ although
the best-fitting dispersion of the log-normal is narrower at $\sigma=0.34$\,dex.

A number of prior and subsequent studies of the Pleiades using similar
methods are consistent with these results. Very similar work has now
been carried out in a number of open clusters with ages of 30--150\,Myr
(Alpha Per, Barrado y Navascu\'es
\etal\ 2002; NGC~2547, Jeffries \etal\ 2004; IC
4665, de Wit \etal\ 2006; Blanco 1, Moraux \etal\ 2007). In common with
the Pleiades studies these clusters are either too old or too distant
for systematic spectroscopic studies of the very low mass and BD
populations and rely mostly on selection in multiple CMDs and in some
cases PMs and a handful of spectra.  
The
results for these clusters also agree with those for the Pleiades; the
IMF below 0.3\msun\ could be represented with a log-normal function with
$m_c \simeq 0.25\msun$ and $\sigma\simeq 0.5$\,dex, or a power law with
$\alpha \simeq 0.5$ (and power laws with larger $\alpha$ at higher
masses), with no evidence for significant variations from
cluster to cluster.

The agreement between these clusters is encouraging, but even in these
relatively young clusters there are clear signs of mass segregation
(Jeffries \etal\ 2004; Moraux \etal\ 2007). As these surveys tend to be
centred on cluster cores, there is a risk that low-mass objects are
under-represented. This is readily apparent in older
clusters that have been analysed in a similar way, where 90 per cent of
BDs may have been evaporated (e.g. the Hyades, age 600\,Myr, Bouvier
\etal\ 2008). On the contrary, other older clusters still agree quite
well with the Pleiades IMF (e.g. Praesepe, age 600\,Myr, Boudreault
\etal\ 2010).  The reasons for these discrepancies are unclear. They
may reflect differing IMFs or they may arise from differing dynamical
histories.  This uncertainty may lead to {\it systematic}
underestimates of the very low-mass IMFs of clusters.

\subsection{Results for younger clusters and star forming regions}

\begin{figure}
\begin{center}
\includegraphics[width=10cm]{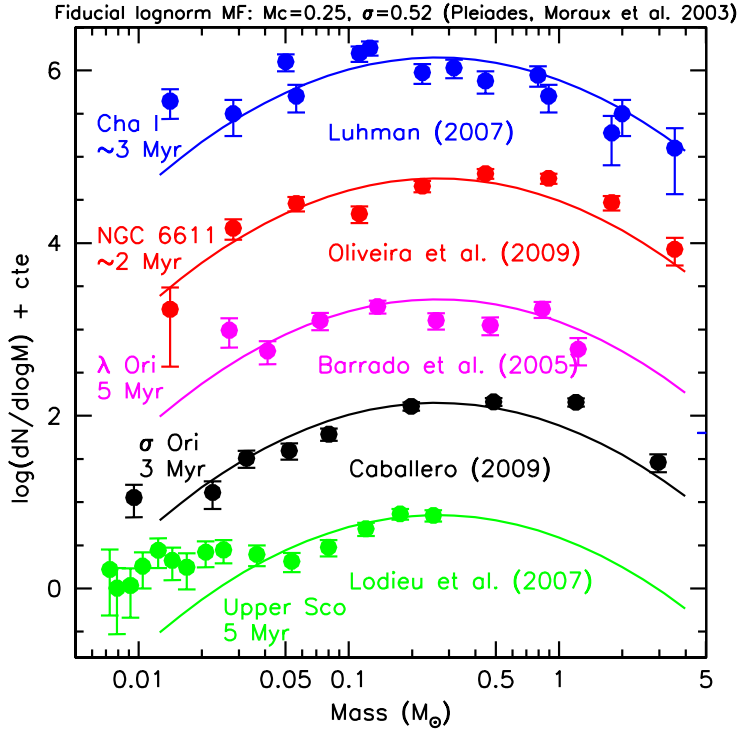}
\end{center}
\caption{The IMF determined in a number of young ($<10$\,Myr) clusters
  and star forming regions (offset for clarity). The solid lines show
  the log-normal model that best fits the Pleiades (see
  Fig.~\ref{yocmf}). The MFs may be generally consistent with that of
  the Pleiades but the MF of Upper Sco is quite different. 
  Figure constructed by Bouvier \& Moraux.
} 
\label{sfrmf}
\end{figure}

The possibilities of dynamical mass segregation drive us to consider
measuring IMFs in very young clusters and star forming regions, with
the additional advantage that very low-mass stars and BDs are much
brighter at younger ages (see Fig.~\ref{kvsm}).  
The presence of extinction, discs and possible age spreads requires a
different approach to measuring IMFs. A
summary of earlier observational work is provided by Luhman \etal\ (2007) and
I consider the determination of the IMF in Chamaeleon~I by Luhman
(2007) as a representative example.

Chamaeleon~I has an age of $\simeq 3$\,Myr and a distance of 170\,pc. An
initial selection of candidate members is made using a number of
generous cuts in far-red and near-infrared CMDs, using the positions of
previously identified members (through H$\alpha$ emission, X-ray
emission and infrared excesses) as a guide. The depths of these
surveys (approximately $I<21$, $H,K<14$ for a wide area and a smaller
region covered to $I<26$) gave access to the IMF over the range
$0.01<m/\msun<3.5$.  Luhman obtained flux-calibrated, low resolution
($600<R<2000$) far-red and near-infrared spectra of the candidates,
using RVs, alkali line strengths, Li absorption and infrared excesses
to confirm membership, determine spectral types and estimate
extinction. Cluster members were placed on an HR diagram using spectral
types to estimate temperatures and bolometric corrections to the
absolute $J$ magnitudes.  Evolutionary tracks are then used to estimate
both the masses and ages of each object. This differs from the
procedure for older clusters because the {\it apparent} spread of age
in Chamaeleon~I of from $<1$ to $\sim 20$\,Myr means that a single
mass-luminosity or mass-magnitude relationship could not apply to all
objects.  Complete, extinction-limited samples are then defined for
each of the photometric surveys and the masses define IMFs for a wider
and an inner region of the cluster respectively.

The two IMFs determined by Luhman (2007) for Chamaeleon~I are
consistent with each other. They are also consistent with an IMF
determined in a similar way for IC~348 (age $\sim 3$\,Myr) by Luhman
\etal\ (2003), and could be represented with a log-normal IMF that peaks
at $m_c \simeq 0.15\msun$. In fact the IMF of a number of young
clusters and star forming regions are remarkably similar (see
Fig.~\ref{sfrmf}), both to each
other and to the IMFs of the Pleiades and the older clusters shown in
Fig.~\ref{yocmf} (Barrado y Navascu\'es \etal\ 2005; Oliveira
\etal\ 2009; Caballero \etal\ 2009). Nevertheless, some clusters do
appear to have a different IMF; Upper Sco has a significant excess of
BDs (Lodieu \etal\ 2007b) compared to other clusters and the IMF of the
Taurus-Auriga association clearly peaks at higher masses (see Luhman
2007).

An alternative metric for comparing the IMFs is the ratio of stars
($0.08$--$1\,\msun$) to BDs ($0.03$--$0.08\,\msun$) (Andersen \etal\
2008; Oliveira \etal\ 2009). This ratio varies from $3.3^{+0.8}_{-0.7}$
in the Orion Nebula cluster (ONC) to $8.3^{+3.3}_{-2.6}$ in
IC\,348. There are no clear trends with cluster size, density or the
presence of strongly ionising O-stars. Andersen \etal\ (2008) present
an analysis that suggests the variations seen in this ratio are
entirely consistent with all the clusters considered (Taurus, Pleiades,
ONC, Mon R2, Chamaeleon~I, NGC~2024 and IC\,348, but they did not
include Upper Sco)
being drawn from the same log-normal IMF proposed by Chabrier (2005).

\subsection{Systematic theoretical uncertainties}

\label{systematics}

\begin{figure}
\begin{center}
\includegraphics[width=10cm]{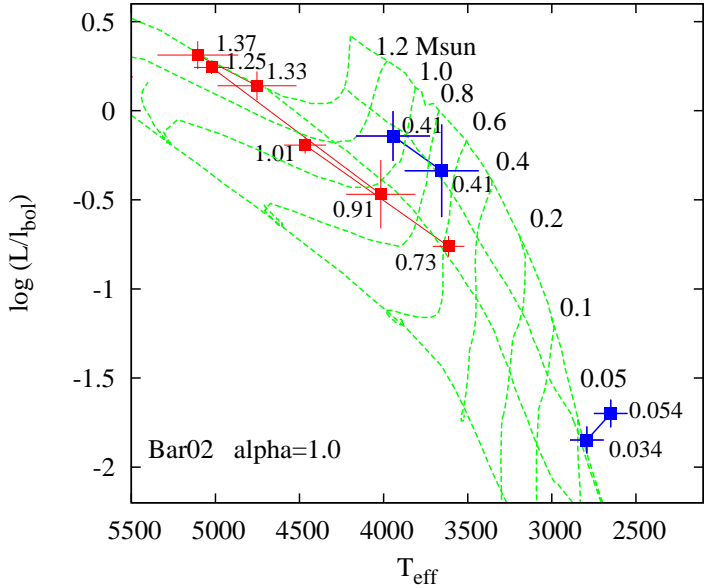}
\end{center}
\caption{Components of pre-main sequence eclipsing binary systems
  placed in the HR diagram (see text). Lines join components of the
  same (and presumably coeval) binary system. The objects are labelled with
  their dynamical masses (in $\msun$) and compared with evolutionary
  tracks (also labelled in $\msun$) from Baraffe \etal\
  (2002). Isochrones are also shown at ages of 1, 3, 10 and
  100\,Myr. The dynamical masses are determined with precisions of 1--8
  per cent and in several cases show significant disagreement with the
  evolutionary tracks. The objects highlighted in blue are Par~1802AB
  and 2M0535-05AB, which are discussed in section~\ref{systematics}.
} 
\label{binaryhr}
\end{figure}

Aside from the problems of dynamical evolution and either ongoing or
primordial mass segregation or ejection, there are a number of
theoretical uncertainties which may have more of an impact on attempts
to determine the IMF in young clusters than in field stars.

The mass-luminosity (or mass-magnitude) relationships are highly
dependent upon the assumed ages of PMS stars and young BDs. But there
is considerable suspicion about the age spreads, or even the absolute
ages, deduced for very young clusters from HR diagrams and model
isochrones (e.g. Naylor 2009). Even the originators of some of the
evolutionary models caution against their reliability for ages
$<10$\,Myr (e.g. Baraffe \etal\ 2002), which is of course the regime where
the majority of the very young clusters are. Unresolved issues include
the treatment of convection and an appropriate representation of the
atmospheres for cool stars where molecules become important. It is well
known that the use of different sets of evolutionary models will lead
to different IMFs in the young clusters, not changing the overall
shape, but significantly changing the fitted parameters of log-normal
parameterisations or power-law exponents (e.g. Da Rio \etal\ 2012).

Beyond these diffiulties, most models also neglect physical effects
that may give rise to significant errors in the determinations of mass
from evolutionary tracks. There is a fierce debate about whether ongoing
or earlier episodes of heavy accretion might lead to a drastic
modification of the luminosity of objects with ages $<10$\,Myr
(e.g. Baraffe \etal\ 2009; Hosokawa \etal\ 2011). This would
systematically change both the deduced ages and masses of stars and BDs
from the HR diagram, but worse, the effect would depend on the {\it
history} of the object rather than its currently observed properties!
At older ages it is likely that the star ``forgets'' its past and
relaxes back towards the standard evolutionary tracks. However,
even for low-mass stars in older clusters at $\sim 100$\,Myr there are strong
suggestions that their strong magnetic fields and starspots could
modify their luminosity and radii, leading to erroneous mass
determinations based on current models that neglect these effects
(e.g. Jackson \etal\ 2009; Mohanty \etal\ 2010).

There has been much less work testing the validity of mass
determinations in PMS stars and BDs than for older field
objects. The main problem is a lack of suitable eclipsing binary
systems and the difficulty of resolving astrometric binary systems at
the distances of the nearest clusters. Those measurements that exist
indicate a large scatter ($\sim 50$ per cent) when comparing dynamical
masses and masses from evolutionary tracks for astrometric binaries
with $0.4 < m/\msun < 1.5$ (Hillenbrand \& White 2004). Among PMS
eclipsing binaries the situation is worse! Figure~\ref{binaryhr} shows
the HR diagram for the components of PMS binaries in Orion, taken
from Covino \etal\ (2001), Stassun \etal\ (2004) and Stempels \etal\
(2008).  The stars are labelled with their
derived dynamical masses and compared to the evolutionary tracks of
Baraffe \etal\ (2002). The positions of some of the binary components
are in agreement with their measured dynamical mass, but others are
very different. Higlighted in blue are two recent awkward examples from
Stassun \etal\ (2006, 2008).
Par~1802AB have equal masses of $(0.41\pm 0.02)\,\msun$
yet masses deduced from the HR diagram would be unequal and twice as
large in the case of one component. In the BD binary system
2M0535-05AB, the more massive component is twice the mass suggested by
evolutionary tracks and cooler than the less massive component.

\subsection{Summary of the IMF from young clusters}

Clusters offer significant advantages (coeval, same composition, known
age and distance, brighter BDs) over field star studies, but also suffer
from the disadvantages of difficult luminosity estimates
and the binary properties of
samples cannot be investigated easily. Whilst significant efforts are
required to exclude contamination and prove cluster membership in
complete samples there is reasonable evidence that most clusters have
IMFs that are consistent with a log-normal representation
(equation~\ref{lognorm}), with $m_c \simeq 0.2\,\msun$ and $\sigma \sim
0.5$\,dex. An equally valid representation of the IMF for $m<0.3\,\msun$ would
be a power-law with $\alpha \simeq 0.5$. Some older clusters show clear
signs of mass segregation and evaporation of low-mass objects and
there are at least two well-studied 
star forming regions that show IMFs that are
clearly inconsistent with the general picture. There is ample evidence from
current observations of PMS binaries that the HR diagram does not yield
reliable masses.  Systematic uncertainties in the IMF due to choice of
evolutionary models or deficiencies in evolutionary models are unlikely
to affect the cluster-to-cluster comparisons (where IMFs have been
detemined using the same models), but may affect any comparison with
the low mass IMF determined from the field.

\section{The bottom of the initial mass function?}

\label{lec5}

The smallest mass with which a BD can form is an important constraint
for formation theories (see Hennebelle's contribution to this
volume). It may be set by the thermodynamics of the primary
fragmentation phase of a collapsing cloud or circumstellar disc --
``the opacity limit''. The lowest possible masses may be in the region
of 0.005--$0.01\,\msun$ (e.g. Low \& Lynden-Bell 1976), but lower values
have been suggested  (0.001--$0.004\,\msun$, Whitworth \& Stamatellos
2006).

\subsection{Measurements in the field}

Evolutionary models suggest that field BDs with spectral types of T8-T9
have $500<T_{\rm eff}<700$\,K and masses of $0.01<m/\msun < 0.03$ for
typical ages of 1--10\,Gyr. To look for lower mass objects requires the
identification of cooler, less-luminous objects, known as Y-dwarfs.
Some examples of such very cool objects have been reported, thanks
largely to 3--5$\mu$m imaging from the Wide-field Infrared Survey
Explorer satellite (WISE, Cushing \etal\ 2011). These objects are redder than
T8 dwarfs in their [3.6]-[4.6] micron colours, have a broad spread of
$J-H$ colours, 
and present spectra with evidence for ammonia absorption.
A preliminary parallax for one of these Y-dwarfs suggests it is 6--7
magnitudes fainter than a T8 dwarf in the $J$ and $H$ bands
(Kirkpatrick \etal\ 2011).

Unfortunately, without an age estimate, the precise masses of these
objects remain uncertain; model-dependent spectroscopic estimates of masses and radii
via $T_{\rm eff}$ and $\log g$ suggest 0.005--0.03\,$\msun$ (Cushing
  \etal\ 2011). Yet, following the same approach described in
section~\ref{lec3} it may be possible to distinguish between different
combinations of IMF and minimum BD mass, through their effects on the
local space density of Y-dwarfs. Kirkpatrick \etal\ (2011) use a
preliminary census from the WISE data to suggest that they can already
rule out IMFs with $\alpha<-1$ or with a minimum BD mass of
$>0.01\,\msun$.

\subsection{Measurements in clusters}

The prospects for finding very low-mass objects in young clusters are
promising. The known age and distance of clusters means that a
(model-dependent) mass can be readily estimated from a
luminosity. Conversely, there is perhaps less confidence in the models to
accurately predict a mass at young ages. The models of Burrows \etal\
(1997) predict that any T-dwarf in a cluster with age $<30$\,Myr will
have $m<0.01\,\msun$.
Very low-mass BDs in young clusters can also be brighter (see
Fig.~\ref{kvsm}). At 5\,Gyr, a $0.005\,\msun$ field object at 5\,pc has
$H\simeq 27$, $T_{\rm eff} \simeq 200$\,K, whereas the same object in
a 5\,Myr old cluster at 400\,pc will have $H \simeq 20$, $T_{\rm eff}
\simeq 1400$\,K (Baraffe \etal\ 2003).

The first claims for the discovery of very low-mass BDs (also
called isolated planetary mass objects or ``planemos'') were made by Lucas \&
Roche (2000) and Zapatero-Osorio \etal\ (2000) in the ONC and
around $\sigma$~Ori respectively. 
They used near-infrared CMDs to select objects and
used their absolute magnitudes to infer masses at or below $0.01\,\msun$. Of
course the possibility of contamination by background or foreground
objects must be ruled out. Spectroscopy on such faint objects ($H>19$)
is very difficult, but Zapater-Osorio \etal\ (2000) and 
Lucas \etal\ (2006) compared some low-resolution
near infrared spectra with those of nearby field BDs and concluded
that their candidates had spectral types cooler than M9. The spectra
suggest a low-gravity but it is unclear at present what the spectra of
very low-gravity L- and T-dwarfs look like.

Another technique aimed specifically at finding T-dwarfs is
``methane-imaging'' (Tinney \etal\ 2005). Narrow-band photometric
filters, approximately 0.12\,$\mu$m wide, are centred at and just
blueward of the deep methane absorption band seen in T-dwarfs at
1.6--1.8\,$\mu$m. The flux ratio (or colour) defined by these bandpasses
becomes very blue when strong methane absorption is established in mid
T-dwarfs and is only weakly affected by reddening.

Attempts to exploit this technique are reported by Burgess \etal\
(2009) in IC~348 ($\sim 3$\,Myr), Haisch \etal\ (2010) around $\rho$~Oph
(age $\sim 1$\,Myr) and Pe{\~n}a Ram{\'{\i}}rez \etal\ (2011) around $\sigma$~Ori (age
$\sim 3$\,Myr.  Burgess \etal\ find 3  candidate T-dwarf members of
IC~348. Estimating their spectral types is complicated by extinction
but also because calibration of the methane narrow band indices must be
done either with older field T-dwarfs or using models that suggest that
young T-dwarfs have a lower $T_{\rm eff}$ for the same methane index.
However two of these candidates are rejected as T-dwarfs by virtue of 
being too blue in $z-J$. Burgess \etal\ show that foreground or
background contamination by field T-dwarfs or extragalactic objects is
unlikely. The nature of these contaminants is therefore unclear. The
remaining candidate could have a mass in the range
0.001--0.005\,\msun. The presence of one object in this mass range is
consistent with an extrapolation of the log-normal IMF that fits the
low-mass stars of the cluster. In contrast, an extrapolation of an
$\alpha=0.6$ power law IMF would predict an order of magnitude more T-dwarfs.

Similar work by Pe\~na Ram{\'{\i}}rez \etal\ (2011) finds 2 T-dwarf
candidates near $\sigma$~Ori. Both have proper motions inconsistent
with cluster membership, yet the probability of contamination by
foreground field T-dwarfs is claimed to be 
very low. Haisch \etal\ (2010) find 22
candidate T-dwarfs in $\rho$~Oph and again, calculate that foreground
and background contamination should be negligible. This number would be
almost 10 per cent of the total cluster population determined by Alves
de Oliveira \etal\ (2012) and would seem more consistent with a
power-law IMF extrapolation than a log-normal IMF. None of these
candidates have so far been confirmed as cluster members with
spectroscopy or PMs, but Alves de Oliveira \etal\ do
spectroscopically confirm 5 L-dwarfs with inferred masses of $\simeq
0.01\,\msun$.

Hence observations in the field and in young clusters suggest that the
bottom of the IMF is not reached until at least $0.01\,\msun$ and there
are a number of candidate objects, which if confirmed, would be of
lower mass. However a problem common to all these studies is
their complete reliance on the fidelity of theoretical models in mass
and age regimes almost untested by empirical data.

\section{A Summary and Future Perspectives}

\label{summary}

In this review I have attempted to describe a range of approaches for
determining the shape of the local stellar and substellar IMF and to
highlight their assumptions and limitations.

The IMF of the Galactic disk stellar population is now reasonably well known
and there is broad agreement between determinations based on a census
of the local population and deeper surveys of field stars over more
limited solid angles. The IMF is Salpeter-like above a solar mass but
becomes flat (when expressed as $\phi(\log m)$) for lower masses down
to $0.1\,\msun$. The IMF for $0.1<m/\msun <2$ could be described in
terms of two power laws with a break at $\simeq 0.5\,\msun$ or in terms
of a log-normal with a broad peak at $\sim 0.25\,\msun$. Improvements in
the IMF of low-mass {\it stars} will certainly come about with the
launch of the Gaia satellite (in 2013?), which will provide accurate
parallaxes for $V\leq 19$ (i.e for objects with an absolute magnitude
of 19 at 10\,pc). This will readily define much larger volume-limited
samples and remove much of the statistical uncertainty that currently
exists below 0.3\,\msun. However, much work will still be required to
assess the contribution of binary systems and separately determine the
``system'' and single-star IMFs. To reduce systematic uncertainties,
it would be much better to determine masses using $K$-band
mass-luminosity relationships and to better calibrate these by finding
and monitoring many more low-mass astrometric binary systems and
by quantitatively assessing the 
variations introduced by metallicity dispersions
in the Galactic disk.

The substellar IMF of field objects is much more
uncertain. It could be described by a power
law ($dN/dm \propto m^{-\alpha}$) with $\alpha \leq 0$ or {\it
possibly} with a log-normal IMF that peaks at $m_c \sim 0.2$ with a
reasonably narrow dispersion ($\sigma \leq 0.5$\,dex), though simulations of
the latter have not yet appeared in the peer-reviewed literature. On
the observational side the T-dwarf (and now Y-dwarf) census is still
quite small, but this will be vastly improved by ongoing ground-based
wide field surveys and the extra sensitivity afforded by
WISE (e.g. Kirkpatrick \etal\ 2012). 
Systematic uncertainties are due to limited knowledge of the binary
properties of field BDs and differences in adopted 
relationships between absolute magnitude
and spectral types.  The former can be attacked with new instruments
like SPIROU (see Artigau in this volume) and the spatial resolution
afforded by the upcoming E-ELT (about 6 mas at 1$\mu$m). The latter
requires a significant effort to measure precise trigonometric
parallaxes for very faint objects beyond the reach of
Gaia and to understand the effects of gravity and metallicity on their
spectra.

A more serious problem is that much of the
available data on BDs in binary systems suggests that their masses are
not well-predicted by the luminosities given by the current generation
of evolutionary models. There is an urgent need to obtain masses for
benchmark BDs in binaries with known age. These could be members of
nearby kinematic moving groups or perhaps companions to white
dwarfs. An alternative would be to find BD binaries in the nearest open
clusters, but these will need the resolution of the E-ELT to measure
their orbits.

The MFs of stars and BDs in young clusters already show remarkable
self-consistency and observations of young clusters 
may be the best method of finding any lower limit
to the IMF. Deeper, wider censuses will monitor the effects of
mass segregation and evaporation; longer baseline PMs based on
digital infrared surveys should prove effective in membership
determination. Confirmation from PMs or spectroscopy becomes
more and more important at lower masses, where the declining MF means there may
be relatively few cluster members among a large number of contaminants. 
The status of the evolutionary models at young ages is even more
precarious than for older BDs. Existing measurements clearly show 
large discrepancies between the dynamical masses and masses
deduced from HR diagrams for both PMS stars and young BDs. More astrometric
and particularly eclipsing binaries must be identified in clusters at a
range of ages to in order to adequately test and refine the models.

It is appealing to agree with Chabrier (2005) that almost all IMF
estimates in the field and clusters are now converging on a universal
log-normal form, with $m_c \simeq 0.25\,\msun$ and $\sigma \simeq 0.5$\,dex.
However, agreement of the {\it system} IMFs might imply a {\it disparity} in
the single/component IMFs, given the age-dependence of mass-luminosity
relationships and the likely evolution of the binary population between
young clusters and the field.  A single log-normal form does represent
most young cluster IMFs well, though there are clear exceptions such as
the Taurus and Upper Sco associations. It may be that these exceptions
lead us to important insights on the star and cluster formation
process, but no clear patterns are yet established. The field IMF {\it
may} also have this form, but this is yet to be demonstrated
convincingly. The currently published simulations favour a power law
IMF with $\alpha \leq 0$, an index that appears to be significantly
smaller than the $\alpha \simeq 0.5$ that could represent very low-mass
stars and BDs in clusters. There is an urgent need to update these
simulations to include the log-normal IMF parameters suggested above,
but also to update the evolutionary models and atmospheres they use and
to make predictions about the gravity distribution of field BDs of a
given spectral type that could be tested by observations.  Given the
systematic problems identified with models and observations in both the
the young clusters and field IMF determinations, it would be remarkable
indeed if any close agreement were found at this stage.

\section*{Acknowledgements}

I would like to thank Corinne Charbonnel, C\'eline
Reyl\'e, Mathias Schultheis and the rest of the local 
organising and scientific committees 
for arranging an exceptional meeting and for
inviting me to be a lecturer. Thanks are due to the Programme National
de Physique Stellaire and the CNRS for their financial support. I would
like to acknowledge the helpful discussions I have had with J\'er\^ome
Bouvier, Adam Burgasser, Ben Burningham and Kelle Cruz whilst
preparing this material.

\nocite{kirkpatrick12}
\nocite{kroupa03}
\nocite{zapatero00}
\nocite{jeffrieslirev}
\nocite{alvesoliveira12}
\nocite{ramirez11}
\nocite{burgess09}
\nocite{haisch10}
\nocite{tinney05}
\nocite{kirkpatrick11}
\nocite{cushing11}
\nocite{low76}
\nocite{whitworth06}
\nocite{covino01}
\nocite{stempels08}
\nocite{stassun08}
\nocite{stassun06}
\nocite{hillenbrand04}
\nocite{mohanty10}
\nocite{jackson09}
\nocite{hosokawa11}
\nocite{dario12}
\nocite{baraffe09}
\nocite{naylor09}
\nocite{andersen08}
\nocite{lodieu07b}
\nocite{luhman03}
\nocite{boudreault10}
\nocite{bouvier08}
\nocite{lodieu07}
\nocite{barrado02}
\nocite{dewit06}
\nocite{moraux07}
\nocite{moraux03}
\nocite{an07}
\nocite{stauffer98}
\nocite{lodieu11b}
\nocite{rebolo96}
\nocite{zapatero02}
\nocite{bouy07}
\nocite{luhman07b}
\nocite{luhman04}
\nocite{valdivielso09}
\nocite{hernandez08}
\nocite{mohanty03}
\nocite{flaccomio10}
\nocite{grosso07}
\nocite{berger10}
\nocite{guedel04}
\nocite{reiners10}
\nocite{reiners12}
\nocite{hartman10}
\nocite{irwin07}
\nocite{reiners12}
\nocite{geller10}
\nocite{frinchaboy08}
\nocite{jeffries06}
\nocite{kenyon05}
\nocite{sacco08}
\nocite{caballero10}
\nocite{rochau10}
\nocite{kraus07}
\nocite{deacon04}
\nocite{hambly99}
\nocite{lodieu11}
\nocite{adams02}
\nocite{umbreit05}
\nocite{reipurth01}
\nocite{proszkow09}
\nocite{jackson10}
\nocite{degrijs08}
\nocite{bonatto11}
\nocite{kroupa93}
\nocite{oliveira09}
\nocite{bejar11}
\nocite{lodieu09}
\nocite{jeffries04}
\nocite{chabrier00}
\nocite{huff06}
\nocite{jeffries11}
\nocite{dario10}
\nocite{hillenbrand97}
\nocite{allison09}
\nocite{demarchi10}
\nocite{baraffe09}
\nocite{baraffe02}
\nocite{burrows01}
\nocite{golimowski04}
\nocite{vrba04}
\nocite{cruz03}
\nocite{knapp04}
\nocite{liu06}
\nocite{marley10}
\nocite{chiu06}
\nocite{metchev08}
\nocite{burningham10}
\nocite{reyle10}
\nocite{burgasser07}
\nocite{reid99}
\nocite{baraffe03}
\nocite{deacon06}
\nocite{allen05}
\nocite{burgasser04}
\nocite{cardoso10}
\nocite{king10}
\nocite{nakajima95}
\nocite{rebolo95}
\nocite{lucas00}
\nocite{lucas06}
\nocite{hennebelle08}
\nocite{padoan04}
\nocite{kroupa02}
\nocite{reid97}
\nocite{duquennoy91}
\nocite{fischer92}
\nocite{allen97}
\nocite{malmquist36}
\nocite{butkevich05}
\nocite{schultheis06}
\nocite{zheng01}
\nocite{martini98}
\nocite{gould97}
\nocite{bochanski10}
\nocite{covey08}
\nocite{tinney93}
\nocite{kroupa97}
\nocite{reid02}
\nocite{hanson79}
\nocite{lutz73}
\nocite{jahreiss97}
\nocite{reid07}
\nocite{henry06}
\nocite{lepine03}
\nocite{cruz07}
\nocite{reid04}
\nocite{gizis02}
\nocite{hawley96}
\nocite{reid95}
\nocite{gliese91}
\nocite{gliese69}
\nocite{gliese57}
\nocite{lepine05}
\nocite{hambly04}
\nocite{luyten57}  
\nocite{luyten79a}
\nocite{luyten79b}
\nocite{luhman05}
\nocite{close05}
\nocite{lane01}
\nocite{konopacky10}
\nocite{dupuy09}
\nocite{bouy03}
\nocite{close03}
\nocite{konopacky10}
\nocite{chabrier97}
\nocite{burrows97}
\nocite{baraffe98}
\nocite{siess00}
\nocite{delfosse00}
\nocite{torres10}
\nocite{morales09}
\nocite{andersen91}
\nocite{segransan00}
\nocite{kraus09}
\nocite{wielen74}
\nocite{luyten68}
\nocite{vanrhijn36}
\nocite{luyten41}
\nocite{salpeter55}
\nocite{millerscalo79}
\nocite{chabrier05}
\nocite{demarchi01}
\nocite{parravano11}
\nocite{luhman07}
\nocite{whitworth07}
\nocite{chabrier03}
\nocite{bastian10}
\nocite{whitworth10}
\nocite{bate05a}
\nocite{bate05b}
\nocite{bate09a}
\nocite{bate09b}
\nocite{schmidt59}
\nocite{kennicutt98}
\nocite{madau96}
\nocite{madau98}
\nocite{gonzalez10}
\nocite{reid00}
\nocite{kapteyn02}
\nocite{luyten23}
\nocite{kuiper42}
\nocite{bruzual03}
\nocite{chabrier02}
\nocite{henry93}
\nocite{xia08}

\bibliographystyle{astron}
\bibliography{roscoff}

\begin{thebibliography}{}

\bibitem[\protect\astroncite{{Adams} et~al.}{2002}]{adams02}
{Adams}, J.~D., {Stauffer}, J.~R., {Skrutskie}, M.~F., {Monet}, D.~G.,
  {Portegies Zwart}, S.~F., {Janes}, K.~A., and {Beichman}, C.~A.: 2002,
\newblock {\em \aj} {\bf 124}, 1570

\bibitem[\protect\astroncite{{Allen}}{2007}]{allen97}
{Allen}, P.~R.: 2007,
\newblock {\em \apj} {\bf 668}, 492

\bibitem[\protect\astroncite{{Allen} et~al.}{2005}]{allen05}
{Allen}, P.~R., {Koerner}, D.~W., {Reid}, I.~N., and {Trilling}, D.~E.: 2005,
\newblock {\em \apj} {\bf 625}, 385

\bibitem[\protect\astroncite{{Allison} et~al.}{2009}]{allison09}
{Allison}, R.~J., {Goodwin}, S.~P., {Parker}, R.~J., {de Grijs}, R., {Portegies
  Zwart}, S.~F., and {Kouwenhoven}, M.~B.~N.: 2009,
\newblock {\em \apjl} {\bf 700}, L99

\bibitem[\protect\astroncite{{Alves de Oliveira}
  et~al.}{2012}]{alvesoliveira12}
{Alves de Oliveira}, C., {Moraux}, E., {Bouvier}, J., and {Bouy}, H.: 2012,
\newblock {\em \aap} {\bf 539}, A151

\bibitem[\protect\astroncite{{An} et~al.}{2007}]{an07}
{An}, D., {Terndrup}, D.~M., {Pinsonneault}, M.~H., {Paulson}, D.~B., {Hanson},
  R.~B., and {Stauffer}, J.~R.: 2007,
\newblock {\em \apj} {\bf 655}, 233

\bibitem[\protect\astroncite{{Andersen}}{1991}]{andersen91}
{Andersen}, J.: 1991,
\newblock {\em \aapr} {\bf 3}, 91

\bibitem[\protect\astroncite{{Andersen} et~al.}{2008}]{andersen08}
{Andersen}, M., {Meyer}, M.~R., {Greissl}, J., and {Aversa}, A.: 2008,
\newblock {\em \apjl} {\bf 683}, L183

\bibitem[\protect\astroncite{{Baraffe} et~al.}{1998}]{baraffe98}
{Baraffe}, I., {Chabrier}, G., {Allard}, F., and {Hauschildt}, P.~H.: 1998,
\newblock {\em \aap} {\bf 337}, 403

\bibitem[\protect\astroncite{{Baraffe} et~al.}{2002}]{baraffe02}
{Baraffe}, I., {Chabrier}, G., {Allard}, F., and {Hauschildt}, P.~H.: 2002,
\newblock {\em \aap} {\bf 382}, 563

\bibitem[\protect\astroncite{{Baraffe} et~al.}{2003}]{baraffe03}
{Baraffe}, I., {Chabrier}, G., {Barman}, T.~S., {Allard}, F., and {Hauschildt},
  P.~H.: 2003,
\newblock {\em \aap} {\bf 402}, 701

\bibitem[\protect\astroncite{{Baraffe} et~al.}{2009}]{baraffe09}
{Baraffe}, I., {Chabrier}, G., and {Gallardo}, J.: 2009,
\newblock {\em \apjl} {\bf 702}, L27

\bibitem[\protect\astroncite{{Barrado y Navascu{\'e}s}
  et~al.}{2002}]{barrado02}
{Barrado y Navascu{\'e}s}, D., {Bouvier}, J., {Stauffer}, J.~R., {Lodieu}, N.,
  and {McCaughrean}, M.~J.: 2002,
\newblock {\em \aap} {\bf 395}, 813

\bibitem[\protect\astroncite{{Bastian} et~al.}{2010}]{bastian10}
{Bastian}, N., {Covey}, K.~R., and {Meyer}, M.~R.: 2010,
\newblock {\em \araa} {\bf 48}, 339

\bibitem[\protect\astroncite{{Bate}}{2005}]{bate05b}
{Bate}, M.~R.: 2005,
\newblock {\em \mnras} {\bf 363}, 363

\bibitem[\protect\astroncite{{Bate}}{2009a}]{bate09a}
{Bate}, M.~R.: 2009a,
\newblock {\em \mnras} {\bf 397}, 232

\bibitem[\protect\astroncite{{Bate}}{2009b}]{bate09b}
{Bate}, M.~R.: 2009b,
\newblock {\em \mnras} {\bf 392}, 1363

\bibitem[\protect\astroncite{{Bate} and {Bonnell}}{2005}]{bate05a}
{Bate}, M.~R. and {Bonnell}, I.~A.: 2005,
\newblock {\em \mnras} {\bf 356}, 1201

\bibitem[\protect\astroncite{{B{\'e}jar} et~al.}{2011}]{bejar11}
{B{\'e}jar}, V.~J.~S., {Zapatero Osorio}, M.~R., {Rebolo}, R., {Caballero},
  J.~A., {Barrado}, D., {Mart{\'{\i}}n}, E.~L., {Mundt}, R., and
  {Bailer-Jones}, C.~A.~L.: 2011,
\newblock {\em \apj} {\bf 743}, 64

\bibitem[\protect\astroncite{{Berger} et~al.}{2010}]{berger10}
{Berger}, E., {Basri}, G., {Fleming}, T.~A., {Giampapa}, M.~S., {Gizis}, J.~E.,
  {Liebert}, J., {Mart{\'{\i}}n}, E., {Phan-Bao}, N., and {Rutledge}, R.~E.:
  2010,
\newblock {\em \apj} {\bf 709}, 332

\bibitem[\protect\astroncite{{Bochanski} et~al.}{2010}]{bochanski10}
{Bochanski}, J.~J., {Hawley}, S.~L., {Covey}, K.~R., {West}, A.~A., {Reid},
  I.~N., {Golimowski}, D.~A., and {Ivezi{\'c}}, {\v Z}.: 2010,
\newblock {\em \aj} {\bf 139}, 2679

\bibitem[\protect\astroncite{{Bonatto} and {Bica}}{2011}]{bonatto11}
{Bonatto}, C. and {Bica}, E.: 2011,
\newblock {\em \mnras} {\bf 415}, 313

\bibitem[\protect\astroncite{{Boudreault} et~al.}{2010}]{boudreault10}
{Boudreault}, S., {Bailer-Jones}, C.~A.~L., {Goldman}, B., {Henning}, T., and
  {Caballero}, J.~A.: 2010,
\newblock {\em \aap} {\bf 510}, A27

\bibitem[\protect\astroncite{{Bouvier} et~al.}{2008}]{bouvier08}
{Bouvier}, J., {Kendall}, T., {Meeus}, G., {Testi}, L., {Moraux}, E.,
  {Stauffer}, J.~R., {James}, D., {Cuillandre}, J.-C., {Irwin}, J.,
  {McCaughrean}, M.~J., {Baraffe}, I., and {Bertin}, E.: 2008,
\newblock {\em \aap} {\bf 481}, 661

\bibitem[\protect\astroncite{{Bouy} et~al.}{2003}]{bouy03}
{Bouy}, H., {Brandner}, W., {Mart{\'{\i}}n}, E.~L., {Delfosse}, X., {Allard},
  F., and {Basri}, G.: 2003,
\newblock {\em \aj} {\bf 126}, 1526

\bibitem[\protect\astroncite{{Bouy} et~al.}{2007}]{bouy07}
{Bouy}, H., {Hu{\'e}lamo}, N., {Mart{\'{\i}}n}, E.~L., {Barrado Y
  Navascu{\'e}s}, D., {Sterzik}, M., and {Pantin}, E.: 2007,
\newblock {\em \aap} {\bf 463}, 641

\bibitem[\protect\astroncite{{Bruzual} and {Charlot}}{2003}]{bruzual03}
{Bruzual}, G. and {Charlot}, S.: 2003,
\newblock {\em \mnras} {\bf 344}, 1000

\bibitem[\protect\astroncite{{Burgasser}}{2004}]{burgasser04}
{Burgasser}, A.~J.: 2004,
\newblock {\em \apjs} {\bf 155}, 191

\bibitem[\protect\astroncite{{Burgasser}}{2007}]{burgasser07}
{Burgasser}, A.~J.: 2007,
\newblock {\em \apj} {\bf 659}, 655

\bibitem[\protect\astroncite{{Burgess} et~al.}{2009}]{burgess09}
{Burgess}, A.~S.~M., {Moraux}, E., {Bouvier}, J., {Marmo}, C., {Albert}, L.,
  and {Bouy}, H.: 2009,
\newblock {\em \aap} {\bf 508}, 823

\bibitem[\protect\astroncite{{Burningham} et~al.}{2010}]{burningham10}
{Burningham}, B., {Pinfield}, D.~J., {Lucas}, P.~W., {Leggett}, S.~K.,
  {Deacon}, N.~R., {Tamura}, M., {Tinney}, C.~G., {Lodieu}, N., {Zhang}, Z.~H.,
  {Huelamo}, N., {Jones}, H.~R.~A., {Murray}, D.~N., {Mortlock}, D.~J.,
  {Patel}, M., {Barrado Y Navascu{\'e}s}, D., {Zapatero Osorio}, M.~R.,
  {Ishii}, M., {Kuzuhara}, M., and {Smart}, R.~L.: 2010,
\newblock {\em \mnras} {\bf 406}, 1885

\bibitem[\protect\astroncite{{Burrows} et~al.}{2001}]{burrows01}
{Burrows}, A., {Hubbard}, W.~B., {Lunine}, J.~I., and {Liebert}, J.: 2001,
\newblock {\em Reviews of Modern Physics} {\bf 73}, 719

\bibitem[\protect\astroncite{{Burrows} et~al.}{1997}]{burrows97}
{Burrows}, A., {Marley}, M., {Hubbard}, W.~B., {Lunine}, J.~I., {Guillot}, T.,
  {Saumon}, D., {Freedman}, R., {Sudarsky}, D., and {Sharp}, C.: 1997,
\newblock {\em \apj} {\bf 491}, 856

\bibitem[\protect\astroncite{{Butkevich} et~al.}{2005}]{butkevich05}
{Butkevich}, A.~G., {Berdyugin}, A.~V., and {Teerikorpi}, P.: 2005,
\newblock {\em \mnras} {\bf 362}, 321

\bibitem[\protect\astroncite{{Caballero}}{2010}]{caballero10}
{Caballero}, J.~A.: 2010,
\newblock {\em \aap} {\bf 514}, A18

\bibitem[\protect\astroncite{{Cardoso} et~al.}{2010}]{cardoso10}
{Cardoso}, C.~V., {McCaughrean}, M.~J., {King}, R.~R., {Close}, L.~M.,
  {Scholz}, R.-D., {Lenzen}, R., {Brandner}, W., {Lodieu}, N., {Zinnecker}, H.,
  {Koehler}, R., and {Konopacky}, Q.~M.: 2010,
\newblock {\em Highlights of Astronomy} {\bf 15}, 761

\bibitem[\protect\astroncite{{Chabrier}}{2002}]{chabrier02}
{Chabrier}, G.: 2002,
\newblock {\em \apj} {\bf 567}, 304

\bibitem[\protect\astroncite{{Chabrier}}{2003}]{chabrier03}
{Chabrier}, G.: 2003,
\newblock {\em \pasp} {\bf 115}, 763

\bibitem[\protect\astroncite{{Chabrier}}{2005}]{chabrier05}
{Chabrier}, G.: 2005,
\newblock in {E.~Corbelli, F.~Palla, \& H.~Zinnecker} (ed.), {\em The Initial
  Mass Function 50 Years Later}, Vol. 327 of {\em Astrophysics and Space
  Science Library}, p.~41

\bibitem[\protect\astroncite{{Chabrier} and {Baraffe}}{1997}]{chabrier97}
{Chabrier}, G. and {Baraffe}, I.: 1997,
\newblock {\em \aap} {\bf 327}, 1039

\bibitem[\protect\astroncite{{Chabrier} et~al.}{2000}]{chabrier00}
{Chabrier}, G., {Baraffe}, I., {Allard}, F., and {Hauschildt}, P.: 2000,
\newblock {\em \apj} {\bf 542}, 464

\bibitem[\protect\astroncite{{Chiu} et~al.}{2006}]{chiu06}
{Chiu}, K., {Fan}, X., {Leggett}, S.~K., {Golimowski}, D.~A., {Zheng}, W.,
  {Geballe}, T.~R., {Schneider}, D.~P., and {Brinkmann}, J.: 2006,
\newblock {\em \aj} {\bf 131}, 2722

\bibitem[\protect\astroncite{{Close} et~al.}{2005}]{close05}
{Close}, L.~M., {Lenzen}, R., {Guirado}, J.~C., {Nielsen}, E.~L., {Mamajek},
  E.~E., {Brandner}, W., {Hartung}, M., {Lidman}, C., and {Biller}, B.: 2005,
\newblock {\em \nat} {\bf 433}, 286

\bibitem[\protect\astroncite{{Close} et~al.}{2003}]{close03}
{Close}, L.~M., {Siegler}, N., {Freed}, M., and {Biller}, B.: 2003,
\newblock {\em \apj} {\bf 587}, 407

\bibitem[\protect\astroncite{{Covey} et~al.}{2008}]{covey08}
{Covey}, K.~R., {Hawley}, S.~L., {Bochanski}, J.~J., {West}, A.~A., {Reid},
  I.~N., {Golimowski}, D.~A., {Davenport}, J.~R.~A., {Henry}, T., {Uomoto}, A.,
  and {Holtzman}, J.~A.: 2008,
\newblock {\em \aj} {\bf 136}, 1778

\bibitem[\protect\astroncite{{Covino} et~al.}{2001}]{covino01}
{Covino}, E., {Melo}, C., {Alcal{\'a}}, J.~M., {Torres}, G., {Fern{\'a}ndez},
  M., {Frasca}, A., and {Paladino}, R.: 2001,
\newblock {\em \aap} {\bf 375}, 130

\bibitem[\protect\astroncite{{Cruz} et~al.}{2007}]{cruz07}
{Cruz}, K.~L., {Reid}, I.~N., {Kirkpatrick}, J.~D., {Burgasser}, A.~J.,
  {Liebert}, J., {Solomon}, A.~R., {Schmidt}, S.~J., {Allen}, P.~R., {Hawley},
  S.~L., and {Covey}, K.~R.: 2007,
\newblock {\em \aj} {\bf 133}, 439

\bibitem[\protect\astroncite{{Cruz} et~al.}{2003}]{cruz03}
{Cruz}, K.~L., {Reid}, I.~N., {Liebert}, J., {Kirkpatrick}, J.~D., and
  {Lowrance}, P.~J.: 2003,
\newblock {\em \aj} {\bf 126}, 2421

\bibitem[\protect\astroncite{{Cushing} et~al.}{2011}]{cushing11}
{Cushing}, M.~C., {Kirkpatrick}, J.~D., {Gelino}, C.~R., {Griffith}, R.~L.,
  {Skrutskie}, M.~F., {Mainzer}, A., {Marsh}, K.~A., {Beichman}, C.~A.,
  {Burgasser}, A.~J., {Prato}, L.~A., {Simcoe}, R.~A., {Marley}, M.~S.,
  {Saumon}, D., {Freedman}, R.~S., {Eisenhardt}, P.~R., and {Wright}, E.~L.:
  2011,
\newblock {\em \apj} {\bf 743}, 50

\bibitem[\protect\astroncite{{Da Rio} et~al.}{2012}]{dario12}
{Da Rio}, N., {Robberto}, M., {Hillenbrand}, L.~A., {Henning}, T., and
  {Stassun}, K.~G.: 2012,
\newblock {\em \apj} {\bf 748}, 14

\bibitem[\protect\astroncite{{Da Rio} et~al.}{2010}]{dario10}
{Da Rio}, N., {Robberto}, M., {Soderblom}, D.~R., {Panagia}, N., {Hillenbrand},
  L.~A., {Palla}, F., and {Stassun}, K.~G.: 2010,
\newblock {\em \apj} {\bf 722}, 1092

\bibitem[\protect\astroncite{{de Grijs} et~al.}{2008}]{degrijs08}
{de Grijs}, R., {Goodwin}, S.~P., {Kouwenhoven}, M.~B.~N., and {Kroupa}, P.:
  2008,
\newblock {\em \aap} {\bf 492}, 685

\bibitem[\protect\astroncite{{de Marchi} and {Paresce}}{2001}]{demarchi01}
{de Marchi}, G. and {Paresce}, F.: 2001,
\newblock in {E.~R.~Schielicke} (ed.), {\em Astronomische Gesellschaft Meeting
  Abstracts}, Vol.~18 of {\em Astronomische Gesellschaft Meeting Abstracts}, p.
  551

\bibitem[\protect\astroncite{{De Marchi} et~al.}{2010}]{demarchi10}
{De Marchi}, G., {Paresce}, F., and {Portegies Zwart}, S.: 2010,
\newblock {\em \apj} {\bf 718}, 105

\bibitem[\protect\astroncite{{de Wit} et~al.}{2006}]{dewit06}
{de Wit}, W.~J., {Bouvier}, J., {Palla}, F., {Cuillandre}, J.-C., {James},
  D.~J., {Kendall}, T.~R., {Lodieu}, N., {McCaughrean}, M.~J., {Moraux}, E.,
  {Randich}, S., and {Testi}, L.: 2006,
\newblock {\em \aap} {\bf 448}, 189

\bibitem[\protect\astroncite{{Deacon} and {Hambly}}{2004}]{deacon04}
{Deacon}, N.~R. and {Hambly}, N.~C.: 2004,
\newblock {\em \aap} {\bf 416}, 125

\bibitem[\protect\astroncite{{Deacon} and {Hambly}}{2006}]{deacon06}
{Deacon}, N.~R. and {Hambly}, N.~C.: 2006,
\newblock {\em \mnras} {\bf 371}, 1722

\bibitem[\protect\astroncite{{Delfosse} et~al.}{2000}]{delfosse00}
{Delfosse}, X., {Forveille}, T., {S{\'e}gransan}, D., {Beuzit}, J.-L., {Udry},
  S., {Perrier}, C., and {Mayor}, M.: 2000,
\newblock {\em \aap} {\bf 364}, 217

\bibitem[\protect\astroncite{{Dupuy} et~al.}{2009}]{dupuy09}
{Dupuy}, T.~J., {Liu}, M.~C., and {Ireland}, M.~J.: 2009,
\newblock {\em \apj} {\bf 699}, 168

\bibitem[\protect\astroncite{{Duquennoy} and {Mayor}}{1991}]{duquennoy91}
{Duquennoy}, A. and {Mayor}, M.: 1991,
\newblock {\em \aap} {\bf 248}, 485

\bibitem[\protect\astroncite{{Fischer} and {Marcy}}{1992}]{fischer92}
{Fischer}, D.~A. and {Marcy}, G.~W.: 1992,
\newblock {\em \apj} {\bf 396}, 178

\bibitem[\protect\astroncite{{Flaccomio} et~al.}{2010}]{flaccomio10}
{Flaccomio}, E., {Micela}, G., {Favata}, F., and {Alencar}, S.~P.~H.: 2010,
\newblock {\em \aap} {\bf 516}, L8

\bibitem[\protect\astroncite{{Frinchaboy} and {Majewski}}{2008}]{frinchaboy08}
{Frinchaboy}, P.~M. and {Majewski}, S.~R.: 2008,
\newblock {\em \aj} {\bf 136}, 118

\bibitem[\protect\astroncite{{Geller} et~al.}{2010}]{geller10}
{Geller}, A.~M., {Mathieu}, R.~D., {Braden}, E.~K., {Meibom}, S., {Platais},
  I., and {Dolan}, C.~J.: 2010,
\newblock {\em \aj} {\bf 139}, 1383

\bibitem[\protect\astroncite{{Gizis} et~al.}{2002}]{gizis02}
{Gizis}, J.~E., {Reid}, I.~N., and {Hawley}, S.~L.: 2002,
\newblock {\em \aj} {\bf 123}, 3356

\bibitem[\protect\astroncite{{Gliese}}{1957}]{gliese57}
{Gliese}, W.: 1957,
\newblock {\em Astron.~Rechen-Institut, Heidelberg, 89 Seiten} {\bf 8}, 1

\bibitem[\protect\astroncite{{Gliese}}{1969}]{gliese69}
{Gliese}, W.: 1969,
\newblock {\em Veroeffentlichungen des Astronomischen Rechen-Instituts
  Heidelberg} {\bf 22}, 1

\bibitem[\protect\astroncite{{Gliese} and {Jahrei{\ss}}}{1991}]{gliese91}
{Gliese}, W. and {Jahrei{\ss}}, H.: 1991,
\newblock {\em {Preliminary Version of the Third Catalogue of Nearby Stars}},
\newblock Technical report

\bibitem[\protect\astroncite{{Golimowski} et~al.}{2004}]{golimowski04}
{Golimowski}, D.~A., {Leggett}, S.~K., {Marley}, M.~S., {Fan}, X., {Geballe},
  T.~R., {Knapp}, G.~R., {Vrba}, F.~J., {Henden}, A.~A., {Luginbuhl}, C.~B.,
  {Guetter}, H.~H., {Munn}, J.~A., {Canzian}, B., {Zheng}, W., {Tsvetanov},
  Z.~I., {Chiu}, K., {Glazebrook}, K., {Hoversten}, E.~A., {Schneider}, D.~P.,
  and {Brinkmann}, J.: 2004,
\newblock {\em \aj} {\bf 127}, 3516

\bibitem[\protect\astroncite{{Gonz{\'a}lez} et~al.}{2010}]{gonzalez10}
{Gonz{\'a}lez}, V., {Labb{\'e}}, I., {Bouwens}, R.~J., {Illingworth}, G.,
  {Franx}, M., {Kriek}, M., and {Brammer}, G.~B.: 2010,
\newblock {\em \apj} {\bf 713}, 115

\bibitem[\protect\astroncite{{Gould} et~al.}{1997}]{gould97}
{Gould}, A., {Bahcall}, J.~N., and {Flynn}, C.: 1997,
\newblock {\em \apj} {\bf 482}, 913

\bibitem[\protect\astroncite{{Grosso} et~al.}{2007}]{grosso07}
{Grosso}, N., {Briggs}, K.~R., {G{\"u}del}, M., {Guieu}, S., {Franciosini}, E.,
  {Palla}, F., {Dougados}, C., {Monin}, J.-L., {M{\'e}nard}, F., {Bouvier}, J.,
  {Audard}, M., and {Telleschi}, A.: 2007,
\newblock {\em \aap} {\bf 468}, 391

\bibitem[\protect\astroncite{{G{\"u}del}}{2004}]{guedel04}
{G{\"u}del}, M.: 2004,
\newblock {\em \aapr} {\bf 12}, 71

\bibitem[\protect\astroncite{{Haisch} et~al.}{2010}]{haisch10}
{Haisch}, Jr., K.~E., {Barsony}, M., and {Tinney}, C.: 2010,
\newblock {\em \apjl} {\bf 719}, L90

\bibitem[\protect\astroncite{{Hambly} et~al.}{2004}]{hambly04}
{Hambly}, N.~C., {Henry}, T.~J., {Subasavage}, J.~P., {Brown}, M.~A., and
  {Jao}, W.-C.: 2004,
\newblock {\em \aj} {\bf 128}, 437

\bibitem[\protect\astroncite{{Hambly} et~al.}{1999}]{hambly99}
{Hambly}, N.~C., {Hodgkin}, S.~T., {Cossburn}, M.~R., and {Jameson}, R.~F.:
  1999,
\newblock {\em \mnras} {\bf 303}, 835

\bibitem[\protect\astroncite{{Hanson}}{1979}]{hanson79}
{Hanson}, R.~B.: 1979,
\newblock {\em \mnras} {\bf 186}, 875

\bibitem[\protect\astroncite{{Hartman} et~al.}{2010}]{hartman10}
{Hartman}, J.~D., {Bakos}, G.~{\'A}., {Kov{\'a}cs}, G., and {Noyes}, R.~W.:
  2010,
\newblock {\em \mnras} {\bf 408}, 475

\bibitem[\protect\astroncite{{Hawley} et~al.}{1996}]{hawley96}
{Hawley}, S.~L., {Gizis}, J.~E., and {Reid}, I.~N.: 1996,
\newblock {\em \aj} {\bf 112}, 2799

\bibitem[\protect\astroncite{{Hennebelle} and {Chabrier}}{2008}]{hennebelle08}
{Hennebelle}, P. and {Chabrier}, G.: 2008,
\newblock {\em \apj} {\bf 684}, 395

\bibitem[\protect\astroncite{{Henry} et~al.}{2006}]{henry06}
{Henry}, T.~J., {Jao}, W.-C., {Subasavage}, J.~P., {Beaulieu}, T.~D., {Ianna},
  P.~A., {Costa}, E., and {M{\'e}ndez}, R.~A.: 2006,
\newblock {\em \aj} {\bf 132}, 2360

\bibitem[\protect\astroncite{{Henry} and {McCarthy}}{1993}]{henry93}
{Henry}, T.~J. and {McCarthy}, Jr., D.~W.: 1993,
\newblock {\em \aj} {\bf 106}, 773

\bibitem[\protect\astroncite{{Hern{\'a}ndez} et~al.}{2008}]{hernandez08}
{Hern{\'a}ndez}, J., {Hartmann}, L., {Calvet}, N., {Jeffries}, R.~D.,
  {Gutermuth}, R., {Muzerolle}, J., and {Stauffer}, J.: 2008,
\newblock {\em \apj} {\bf 686}, 1195

\bibitem[\protect\astroncite{{Hillenbrand}}{1997}]{hillenbrand97}
{Hillenbrand}, L.~A.: 1997,
\newblock {\em \aj} {\bf 113}, 1733

\bibitem[\protect\astroncite{{Hillenbrand} and {White}}{2004}]{hillenbrand04}
{Hillenbrand}, L.~A. and {White}, R.~J.: 2004,
\newblock {\em \apj} {\bf 604}, 741

\bibitem[\protect\astroncite{{Hosokawa} et~al.}{2011}]{hosokawa11}
{Hosokawa}, T., {Offner}, S.~S.~R., and {Krumholz}, M.~R.: 2011,
\newblock {\em \apj} {\bf 738}, 140

\bibitem[\protect\astroncite{{Huff} and {Stahler}}{2006}]{huff06}
{Huff}, E.~M. and {Stahler}, S.~W.: 2006,
\newblock {\em \apj} {\bf 644}, 355

\bibitem[\protect\astroncite{{Irwin} et~al.}{2007}]{irwin07}
{Irwin}, J., {Hodgkin}, S., {Aigrain}, S., {Hebb}, L., {Bouvier}, J., {Clarke},
  C., {Moraux}, E., and {Bramich}, D.~M.: 2007,
\newblock {\em \mnras} {\bf 377}, 741

\bibitem[\protect\astroncite{{Jackson} and {Jeffries}}{2010}]{jackson10}
{Jackson}, R.~J. and {Jeffries}, R.~D.: 2010,
\newblock {\em \mnras} {\bf 407}, 465

\bibitem[\protect\astroncite{{Jackson} et~al.}{2009}]{jackson09}
{Jackson}, R.~J., {Jeffries}, R.~D., and {Maxted}, P.~F.~L.: 2009,
\newblock {\em \mnras} {\bf 399}, L89

\bibitem[\protect\astroncite{{Jahrei{\ss}} and {Wielen}}{1997}]{jahreiss97}
{Jahrei{\ss}}, H. and {Wielen}, R.: 1997,
\newblock in {R.~M.~Bonnet, E.~H{\o}g, P.~L.~Bernacca, L.~Emiliani, A.~Blaauw,
  C.~Turon, J.~Kovalevsky, L.~Lindegren, H.~Hassan, M.~Bouffard, B.~Strim,
  D.~Heger, M.~A.~C.~Perryman, \& L.~Woltjer} (ed.), {\em Hipparcos - Venice
  '97}, Vol. 402 of {\em ESA Special Publication}, pp 675--680

\bibitem[\protect\astroncite{{Jeffries}}{2006}]{jeffrieslirev}
{Jeffries}, R.~D.: 2006,
\newblock in S. {Randich} and L. {Pasquini} (eds.), {\em Chemical Abundances
  and Mixing in Stars in the Milky Way and its Satellites}, ESO Astrophysics
  Symposia, p. 163

\bibitem[\protect\astroncite{{Jeffries} et~al.}{2011}]{jeffries11}
{Jeffries}, R.~D., {Littlefair}, S.~P., {Naylor}, T., and {Mayne}, N.~J.: 2011,
\newblock {\em \mnras} {\bf 418}, 1948

\bibitem[\protect\astroncite{{Jeffries} et~al.}{2006}]{jeffries06}
{Jeffries}, R.~D., {Maxted}, P.~F.~L., {Oliveira}, J.~M., and {Naylor}, T.:
  2006,
\newblock {\em \mnras} {\bf 371}, L6

\bibitem[\protect\astroncite{{Jeffries} et~al.}{2004}]{jeffries04}
{Jeffries}, R.~D., {Naylor}, T., {Devey}, C.~R., and {Totten}, E.~J.: 2004,
\newblock {\em \mnras} {\bf 351}, 1401

\bibitem[\protect\astroncite{{Kapteyn}}{1902}]{kapteyn02}
{Kapteyn}, J.~C.: 1902,
\newblock {\em Publications of the Kapteyn Astronomical Laboratory Groningen}
  {\bf 11}, 3

\bibitem[\protect\astroncite{{Kennicutt}}{1998}]{kennicutt98}
{Kennicutt}, Jr., R.~C.: 1998,
\newblock {\em \apj} {\bf 498}, 541

\bibitem[\protect\astroncite{{Kenyon} et~al.}{2005}]{kenyon05}
{Kenyon}, M.~J., {Jeffries}, R.~D., {Naylor}, T., {Oliveira}, J.~M., and
  {Maxted}, P.~F.~L.: 2005,
\newblock {\em \mnras} {\bf 356}, 89

\bibitem[\protect\astroncite{{King} et~al.}{2010}]{king10}
{King}, R.~R., {McCaughrean}, M.~J., {Homeier}, D., {Allard}, F., {Scholz},
  R.-D., and {Lodieu}, N.: 2010,
\newblock {\em \aap} {\bf 510}, A99

\bibitem[\protect\astroncite{{Kirkpatrick} et~al.}{2011}]{kirkpatrick11}
{Kirkpatrick}, J.~D., {Cushing}, M.~C., {Gelino}, C.~R., {Griffith}, R.~L.,
  {Skrutskie}, M.~F., {Marsh}, K.~A., {Wright}, E.~L., {Mainzer}, A.,
  {Eisenhardt}, P.~R., {McLean}, I.~S., {Thompson}, M.~A., {Bauer}, J.~M.,
  {Benford}, D.~J., {Bridge}, C.~R., {Lake}, S.~E., {Petty}, S.~M., {Stanford},
  S.~A., {Tsai}, C.-W., {Bailey}, V., {Beichman}, C.~A., {Bloom}, J.~S.,
  {Bochanski}, J.~J., {Burgasser}, A.~J., {Capak}, P.~L., {Cruz}, K.~L.,
  {Hinz}, P.~M., {Kartaltepe}, J.~S., {Knox}, R.~P., {Manohar}, S., {Masters},
  D., {Morales-Calder{\'o}n}, M., {Prato}, L.~A., {Rodigas}, T.~J., {Salvato},
  M., {Schurr}, S.~D., {Scoville}, N.~Z., {Simcoe}, R.~A., {Stapelfeldt},
  K.~R., {Stern}, D., {Stock}, N.~D., and {Vacca}, W.~D.: 2011,
\newblock {\em \apjs} {\bf 197}, 19

\bibitem[\protect\astroncite{{Kirkpatrick} et~al.}{2012}]{kirkpatrick12}
{Kirkpatrick}, J.~D., {Gelino}, C.~R., {Cushing}, M.~C., {Mace}, G.~N.,
  {Griffith}, R.~L., {Skrutskie}, M.~F., {Marsh}, K.~A., {Wright}, E.~L.,
  {Eisenhardt}, P.~R., {McLean}, I.~S., {Mainzer}, A.~K., {Burgasser}, A.~J.,
  {Tinney}, C.~G., {Parker}, S., and {Salter}, G.: 2012,
\newblock {\em ArXiv e-prints, 1205.2122}

\bibitem[\protect\astroncite{{Knapp} et~al.}{2004}]{knapp04}
{Knapp}, G.~R., {Leggett}, S.~K., {Fan}, X., {Marley}, M.~S., {Geballe}, T.~R.,
  {Golimowski}, D.~A., {Finkbeiner}, D., {Gunn}, J.~E., {Hennawi}, J.,
  {Ivezi{\'c}}, Z., {Lupton}, R.~H., {Schlegel}, D.~J., {Strauss}, M.~A.,
  {Tsvetanov}, Z.~I., {Chiu}, K., {Hoversten}, E.~A., {Glazebrook}, K.,
  {Zheng}, W., {Hendrickson}, M., {Williams}, C.~C., {Uomoto}, A., {Vrba},
  F.~J., {Henden}, A.~A., {Luginbuhl}, C.~B., {Guetter}, H.~H., {Munn}, J.~A.,
  {Canzian}, B., {Schneider}, D.~P., and {Brinkmann}, J.: 2004,
\newblock {\em \aj} {\bf 127}, 3553

\bibitem[\protect\astroncite{{Konopacky} et~al.}{2010}]{konopacky10}
{Konopacky}, Q.~M., {Ghez}, A.~M., {Barman}, T.~S., {Rice}, E.~L., {Bailey},
  III, J.~I., {White}, R.~J., {McLean}, I.~S., and {Duch{\^e}ne}, G.: 2010,
\newblock {\em \apj} {\bf 711}, 1087

\bibitem[\protect\astroncite{{Kraus} and {Hillenbrand}}{2007}]{kraus07}
{Kraus}, A.~L. and {Hillenbrand}, L.~A.: 2007,
\newblock {\em \aj} {\bf 134}, 2340

\bibitem[\protect\astroncite{{Kraus} et~al.}{2009}]{kraus09}
{Kraus}, S., {Weigelt}, G., {Balega}, Y.~Y., {Docobo}, J.~A., {Hofmann}, K.-H.,
  {Preibisch}, T., {Schertl}, D., {Tamazian}, V.~S., {Driebe}, T., {Ohnaka},
  K., {Petrov}, R., {Sch{\"o}ller}, M., and {Smith}, M.: 2009,
\newblock {\em \aap} {\bf 497}, 195

\bibitem[\protect\astroncite{{Kroupa}}{2002}]{kroupa02}
{Kroupa}, P.: 2002,
\newblock {\em Science} {\bf 295}, 82

\bibitem[\protect\astroncite{{Kroupa} and {Bouvier}}{2003}]{kroupa03}
{Kroupa}, P. and {Bouvier}, J.: 2003,
\newblock {\em \mnras} {\bf 346}, 343

\bibitem[\protect\astroncite{{Kroupa} and {Tout}}{1997}]{kroupa97}
{Kroupa}, P. and {Tout}, C.~A.: 1997,
\newblock {\em \mnras} {\bf 287}, 402

\bibitem[\protect\astroncite{{Kroupa} et~al.}{1993}]{kroupa93}
{Kroupa}, P., {Tout}, C.~A., and {Gilmore}, G.: 1993,
\newblock {\em \mnras} {\bf 262}, 545

\bibitem[\protect\astroncite{{Kuiper}}{1942}]{kuiper42}
{Kuiper}, G.~P.: 1942,
\newblock {\em \apj} {\bf 95}, 201

\bibitem[\protect\astroncite{{Lane} et~al.}{2001}]{lane01}
{Lane}, B.~F., {Zapatero Osorio}, M.~R., {Britton}, M.~C., {Mart{\'{\i}}n},
  E.~L., and {Kulkarni}, S.~R.: 2001,
\newblock {\em \apj} {\bf 560}, 390

\bibitem[\protect\astroncite{{L{\'e}pine}}{2005}]{lepine05}
{L{\'e}pine}, S.: 2005,
\newblock {\em \aj} {\bf 130}, 1247

\bibitem[\protect\astroncite{{L{\'e}pine} et~al.}{2003}]{lepine03}
{L{\'e}pine}, S., {Rich}, R.~M., and {Shara}, M.~M.: 2003,
\newblock {\em \aj} {\bf 125}, 1598

\bibitem[\protect\astroncite{{Liu} et~al.}{2006}]{liu06}
{Liu}, M.~C., {Leggett}, S.~K., {Golimowski}, D.~A., {Chiu}, K., {Fan}, X.,
  {Geballe}, T.~R., {Schneider}, D.~P., and {Brinkmann}, J.: 2006,
\newblock {\em \apj} {\bf 647}, 1393

\bibitem[\protect\astroncite{{Lodieu}}{2011}]{lodieu11}
{Lodieu}, N.: 2011,
\newblock {\em Research, Science and Technology of Brown Dwarfs and Exoplanets:
  Proceedings of an International Conference held in Shangai on Occasion of a
  Total Eclipse of the Sun, Shangai, China, Edited by E.L.~Martin; J.~Ge;
  W.~Lin; EPJ Web of Conferences, Volume 16, id.06001} {\bf 16}, 6001

\bibitem[\protect\astroncite{{Lodieu} et~al.}{2007a}]{lodieu07}
{Lodieu}, N., {Dobbie}, P.~D., {Deacon}, N.~R., {Hodgkin}, S.~T., {Hambly},
  N.~C., and {Jameson}, R.~F.: 2007a,
\newblock {\em \mnras} {\bf 380}, 712

\bibitem[\protect\astroncite{{Lodieu} et~al.}{2011}]{lodieu11b}
{Lodieu}, N., {Dobbie}, P.~D., and {Hambly}, N.~C.: 2011,
\newblock {\em \aap} {\bf 527}, A24

\bibitem[\protect\astroncite{{Lodieu} et~al.}{2007b}]{lodieu07b}
{Lodieu}, N., {Hambly}, N.~C., {Jameson}, R.~F., {Hodgkin}, S.~T., {Carraro},
  G., and {Kendall}, T.~R.: 2007b,
\newblock {\em \mnras} {\bf 374}, 372

\bibitem[\protect\astroncite{{Lodieu} et~al.}{2009}]{lodieu09}
{Lodieu}, N., {Zapatero Osorio}, M.~R., {Rebolo}, R., {Mart{\'{\i}}n}, E.~L.,
  and {Hambly}, N.~C.: 2009,
\newblock {\em \aap} {\bf 505}, 1115

\bibitem[\protect\astroncite{{Low} and {Lynden-Bell}}{1976}]{low76}
{Low}, C. and {Lynden-Bell}, D.: 1976,
\newblock {\em \mnras} {\bf 176}, 367

\bibitem[\protect\astroncite{{Lucas} and {Roche}}{2000}]{lucas00}
{Lucas}, P.~W. and {Roche}, P.~F.: 2000,
\newblock {\em \mnras} {\bf 314}, 858

\bibitem[\protect\astroncite{{Lucas} et~al.}{2006}]{lucas06}
{Lucas}, P.~W., {Weights}, D.~J., {Roche}, P.~F., and {Riddick}, F.~C.: 2006,
\newblock {\em \mnras} {\bf 373}, L60

\bibitem[\protect\astroncite{{Luhman}}{2004}]{luhman04}
{Luhman}, K.~L.: 2004,
\newblock {\em \apj} {\bf 617}, 1216

\bibitem[\protect\astroncite{{Luhman}}{2007}]{luhman07b}
{Luhman}, K.~L.: 2007,
\newblock {\em \apjs} {\bf 173}, 104

\bibitem[\protect\astroncite{{Luhman} et~al.}{2007}]{luhman07}
{Luhman}, K.~L., {Joergens}, V., {Lada}, C., {Muzerolle}, J., {Pascucci}, I.,
  and {White}, R.: 2007,
\newblock {\em Protostars and Planets V} pp 443--457

\bibitem[\protect\astroncite{{Luhman} et~al.}{2005}]{luhman05}
{Luhman}, K.~L., {Stauffer}, J.~R., and {Mamajek}, E.~E.: 2005,
\newblock {\em \apjl} {\bf 628}, L69

\bibitem[\protect\astroncite{{Luhman} et~al.}{2003}]{luhman03}
{Luhman}, K.~L., {Stauffer}, J.~R., {Muench}, A.~A., {Rieke}, G.~H., {Lada},
  E.~A., {Bouvier}, J., and {Lada}, C.~J.: 2003,
\newblock {\em \apj} {\bf 593}, 1093

\bibitem[\protect\astroncite{{Lutz} and {Kelker}}{1973}]{lutz73}
{Lutz}, T.~E. and {Kelker}, D.~H.: 1973,
\newblock {\em \pasp} {\bf 85}, 573

\bibitem[\protect\astroncite{{Luyten}}{1923}]{luyten23}
{Luyten}, W.~J.: 1923,
\newblock {\em \pasp} {\bf 35}, 209

\bibitem[\protect\astroncite{{Luyten}}{1941}]{luyten41}
{Luyten}, W.~J.: 1941,
\newblock {\em Annals of the New York Academy of Sciences} {\bf 42}, 201

\bibitem[\protect\astroncite{{Luyten}}{1957}]{luyten57}
{Luyten}, W.~J.: 1957,
\newblock {\em {A catalogue of 9867 stars in the Southern Hemisphere with
  proper motions exceeding 0.''2 annually.}}

\bibitem[\protect\astroncite{{Luyten}}{1968}]{luyten68}
{Luyten}, W.~J.: 1968,
\newblock {\em \mnras} {\bf 139}, 221

\bibitem[\protect\astroncite{{Luyten}}{1979a}]{luyten79a}
{Luyten}, W.~J.: 1979a,
\newblock {\em {LHS catalogue. A catalogue of stars with proper motions
  exceeding 0''5 annually}}

\bibitem[\protect\astroncite{{Luyten}}{1979b}]{luyten79b}
{Luyten}, W.~J.: 1979b,
\newblock {\em {NLTT catalogue. Volume\_I. +90\_\_to\_+30\_. Volume.\_II.
  +30\_\_to\_0\_.}}

\bibitem[\protect\astroncite{{Madau} et~al.}{1996}]{madau96}
{Madau}, P., {Ferguson}, H.~C., {Dickinson}, M.~E., {Giavalisco}, M.,
  {Steidel}, C.~C., and {Fruchter}, A.: 1996,
\newblock {\em \mnras} {\bf 283}, 1388

\bibitem[\protect\astroncite{{Madau} et~al.}{1998}]{madau98}
{Madau}, P., {Pozzetti}, L., and {Dickinson}, M.: 1998,
\newblock {\em \apj} {\bf 498}, 106

\bibitem[\protect\astroncite{{Malmquist}}{1936}]{malmquist36}
{Malmquist}, K.~G.: 1936,
\newblock {\em Stockholms Obs. Medd.} 26

\bibitem[\protect\astroncite{{Marley} et~al.}{2010}]{marley10}
{Marley}, M.~S., {Saumon}, D., and {Goldblatt}, C.: 2010,
\newblock {\em \apjl} {\bf 723}, L117

\bibitem[\protect\astroncite{{Martini} and {Osmer}}{1998}]{martini98}
{Martini}, P. and {Osmer}, P.~S.: 1998,
\newblock {\em \aj} {\bf 116}, 2513

\bibitem[\protect\astroncite{{Metchev} et~al.}{2008}]{metchev08}
{Metchev}, S.~A., {Kirkpatrick}, J.~D., {Berriman}, G.~B., and {Looper}, D.:
  2008,
\newblock {\em \apj} {\bf 676}, 1281

\bibitem[\protect\astroncite{{Miller} and {Scalo}}{1979}]{millerscalo79}
{Miller}, G.~E. and {Scalo}, J.~M.: 1979,
\newblock {\em \apjs} {\bf 41}, 513

\bibitem[\protect\astroncite{{Mohanty} and {Basri}}{2003}]{mohanty03}
{Mohanty}, S. and {Basri}, G.: 2003,
\newblock {\em \apj} {\bf 583}, 451

\bibitem[\protect\astroncite{{Mohanty} et~al.}{2010}]{mohanty10}
{Mohanty}, S., {Stassun}, K.~G., and {Doppmann}, G.~W.: 2010,
\newblock {\em \apj} {\bf 722}, 1138

\bibitem[\protect\astroncite{{Morales} et~al.}{2009}]{morales09}
{Morales}, J.~C., {Ribas}, I., {Jordi}, C., {Torres}, G., {Gallardo}, J.,
  {Guinan}, E.~F., {Charbonneau}, D., {Wolf}, M., {Latham}, D.~W.,
  {Anglada-Escud{\'e}}, G., {Bradstreet}, D.~H., {Everett}, M.~E., {O'Donovan},
  F.~T., {Mandushev}, G., and {Mathieu}, R.~D.: 2009,
\newblock {\em \apj} {\bf 691}, 1400

\bibitem[\protect\astroncite{{Moraux} et~al.}{2007}]{moraux07}
{Moraux}, E., {Bouvier}, J., {Stauffer}, J.~R., {Barrado y Navascu{\'e}s}, D.,
  and {Cuillandre}, J.-C.: 2007,
\newblock {\em \aap} {\bf 471}, 499

\bibitem[\protect\astroncite{{Moraux} et~al.}{2003}]{moraux03}
{Moraux}, E., {Bouvier}, J., {Stauffer}, J.~R., and {Cuillandre}, J.-C.: 2003,
\newblock {\em \aap} {\bf 400}, 891

\bibitem[\protect\astroncite{{Nakajima} et~al.}{1995}]{nakajima95}
{Nakajima}, T., {Oppenheimer}, B.~R., {Kulkarni}, S.~R., {Golimowski}, D.~A.,
  {Matthews}, K., and {Durrance}, S.~T.: 1995,
\newblock {\em \nat} {\bf 378}, 463

\bibitem[\protect\astroncite{{Naylor}}{2009}]{naylor09}
{Naylor}, T.: 2009,
\newblock {\em \mnras} {\bf 399}, 432

\bibitem[\protect\astroncite{{Oliveira} et~al.}{2009}]{oliveira09}
{Oliveira}, J.~M., {Jeffries}, R.~D., and {van Loon}, J.~T.: 2009,
\newblock {\em \mnras} {\bf 392}, 1034

\bibitem[\protect\astroncite{{Padoan} and {Nordlund}}{2004}]{padoan04}
{Padoan}, P. and {Nordlund}, {\AA}.: 2004,
\newblock {\em \apj} {\bf 617}, 559

\bibitem[\protect\astroncite{{Parravano} et~al.}{2011}]{parravano11}
{Parravano}, A., {McKee}, C.~F., and {Hollenbach}, D.~J.: 2011,
\newblock {\em \apj} {\bf 726}, 27

\bibitem[\protect\astroncite{{Pe{\~n}a Ram{\'{\i}}rez}
  et~al.}{2011}]{ramirez11}
{Pe{\~n}a Ram{\'{\i}}rez}, K., {Zapatero Osorio}, M.~R., {B{\'e}jar}, V.~J.~S.,
  {Rebolo}, R., and {Bihain}, G.: 2011,
\newblock {\em \aap} {\bf 532}, A42

\bibitem[\protect\astroncite{{Proszkow} et~al.}{2009}]{proszkow09}
{Proszkow}, E.-M., {Adams}, F.~C., {Hartmann}, L.~W., and {Tobin}, J.~J.: 2009,
\newblock {\em \apj} {\bf 697}, 1020

\bibitem[\protect\astroncite{{Rebolo} et~al.}{1996}]{rebolo96}
{Rebolo}, R., {Martin}, E.~L., {Basri}, G., {Marcy}, G.~W., and
  {Zapatero-Osorio}, M.~R.: 1996,
\newblock {\em \apjl} {\bf 469}, L53

\bibitem[\protect\astroncite{{Rebolo} et~al.}{1995}]{rebolo95}
{Rebolo}, R., {Zapatero Osorio}, M.~R., and {Mart{\'{\i}}n}, E.~L.: 1995,
\newblock {\em \nat} {\bf 377}, 129

\bibitem[\protect\astroncite{{Reid} et~al.}{2004}]{reid04}
{Reid}, I.~N., {Cruz}, K.~L., {Allen}, P., {Mungall}, F., {Kilkenny}, D.,
  {Liebert}, J., {Hawley}, S.~L., {Fraser}, O.~J., {Covey}, K.~R., {Lowrance},
  P., {Kirkpatrick}, J.~D., and {Burgasser}, A.~J.: 2004,
\newblock {\em \aj} {\bf 128}, 463

\bibitem[\protect\astroncite{{Reid} et~al.}{2007}]{reid07}
{Reid}, I.~N., {Cruz}, K.~L., and {Allen}, P.~R.: 2007,
\newblock {\em \aj} {\bf 133}, 2825

\bibitem[\protect\astroncite{{Reid} and {Gizis}}{1997}]{reid97}
{Reid}, I.~N. and {Gizis}, J.~E.: 1997,
\newblock {\em \aj} {\bf 113}, 2246

\bibitem[\protect\astroncite{{Reid} et~al.}{2002}]{reid02}
{Reid}, I.~N., {Gizis}, J.~E., and {Hawley}, S.~L.: 2002,
\newblock {\em \aj} {\bf 124}, 2721

\bibitem[\protect\astroncite{{Reid} and {Hawley}}{2000}]{reid00}
{Reid}, I.~N. and {Hawley}, S.~L.: 2000,
\newblock {\em {New light on dark stars. Red dwarfs, low-mass stars, brown
  dwarfs.}},
\newblock Praxis Publishing, Chichester (UK)

\bibitem[\protect\astroncite{{Reid} et~al.}{1995}]{reid95}
{Reid}, I.~N., {Hawley}, S.~L., and {Gizis}, J.~E.: 1995,
\newblock {\em \aj} {\bf 110}, 1838

\bibitem[\protect\astroncite{{Reid} et~al.}{1999}]{reid99}
{Reid}, I.~N., {Kirkpatrick}, J.~D., {Liebert}, J., {Burrows}, A., {Gizis},
  J.~E., {Burgasser}, A., {Dahn}, C.~C., {Monet}, D., {Cutri}, R., {Beichman},
  C.~A., and {Skrutskie}, M.: 1999,
\newblock {\em \apj} {\bf 521}, 613

\bibitem[\protect\astroncite{{Reiners} and {Basri}}{2010}]{reiners10}
{Reiners}, A. and {Basri}, G.: 2010,
\newblock {\em \apj} {\bf 710}, 924

\bibitem[\protect\astroncite{{Reiners} et~al.}{2012}]{reiners12}
{Reiners}, A., {Joshi}, N., and {Goldman}, B.: 2012,
\newblock {\em \aj} {\bf 143}, 93

\bibitem[\protect\astroncite{{Reipurth} and {Clarke}}{2001}]{reipurth01}
{Reipurth}, B. and {Clarke}, C.: 2001,
\newblock {\em \aj} {\bf 122}, 432

\bibitem[\protect\astroncite{{Reyl{\'e}} et~al.}{2010}]{reyle10}
{Reyl{\'e}}, C., {Delorme}, P., {Willott}, C.~J., {Albert}, L., {Delfosse}, X.,
  {Forveille}, T., {Artigau}, E., {Malo}, L., {Hill}, G.~J., and {Doyon}, R.:
  2010,
\newblock {\em \aap} {\bf 522}, A112

\bibitem[\protect\astroncite{{Rochau} et~al.}{2010}]{rochau10}
{Rochau}, B., {Brandner}, W., {Stolte}, A., {Gennaro}, M., {Gouliermis}, D.,
  {Da Rio}, N., {Dzyurkevich}, N., and {Henning}, T.: 2010,
\newblock {\em \apjl} {\bf 716}, L90

\bibitem[\protect\astroncite{{Sacco} et~al.}{2008}]{sacco08}
{Sacco}, G.~G., {Franciosini}, E., {Randich}, S., and {Pallavicini}, R.: 2008,
\newblock {\em \aap} {\bf 488}, 167

\bibitem[\protect\astroncite{{Salpeter}}{1955}]{salpeter55}
{Salpeter}, E.~E.: 1955,
\newblock {\em \apj} {\bf 121}, 161

\bibitem[\protect\astroncite{{Schmidt}}{1959}]{schmidt59}
{Schmidt}, M.: 1959,
\newblock {\em \apj} {\bf 129}, 243

\bibitem[\protect\astroncite{{Schultheis} et~al.}{2006}]{schultheis06}
{Schultheis}, M., {Robin}, A.~C., {Reyl{\'e}}, C., {McCracken}, H.~J.,
  {Bertin}, E., {Mellier}, Y., and {Le F{\`e}vre}, O.: 2006,
\newblock {\em \aap} {\bf 447}, 185

\bibitem[\protect\astroncite{{S{\'e}gransan} et~al.}{2000}]{segransan00}
{S{\'e}gransan}, D., {Delfosse}, X., {Forveille}, T., {Beuzit}, J.-L., {Udry},
  S., {Perrier}, C., and {Mayor}, M.: 2000,
\newblock {\em \aap} {\bf 364}, 665

\bibitem[\protect\astroncite{{Siess} et~al.}{2000}]{siess00}
{Siess}, L., {Dufour}, E., and {Forestini}, M.: 2000,
\newblock {\em \aap} {\bf 358}, 593

\bibitem[\protect\astroncite{{Stassun} et~al.}{2008}]{stassun08}
{Stassun}, K.~G., {Mathieu}, R.~D., {Cargile}, P.~A., {Aarnio}, A.~N.,
  {Stempels}, E., and {Geller}, A.: 2008,
\newblock {\em \nat} {\bf 453}, 1079

\bibitem[\protect\astroncite{{Stassun} et~al.}{2006}]{stassun06}
{Stassun}, K.~G., {Mathieu}, R.~D., and {Valenti}, J.~A.: 2006,
\newblock {\em \nat} {\bf 440}, 311

\bibitem[\protect\astroncite{{Stauffer} et~al.}{1998}]{stauffer98}
{Stauffer}, J.~R., {Schultz}, G., and {Kirkpatrick}, J.~D.: 1998,
\newblock {\em \apjl} {\bf 499}, L199

\bibitem[\protect\astroncite{{Stempels} et~al.}{2008}]{stempels08}
{Stempels}, H.~C., {Hebb}, L., {Stassun}, K.~G., {Holtzman}, J., {Dunstone},
  N., {Glowienka}, L., and {Frandsen}, S.: 2008,
\newblock {\em \aap} {\bf 481}, 747

\bibitem[\protect\astroncite{{Tinney}}{1993}]{tinney93}
{Tinney}, C.~G.: 1993,
\newblock {\em \apj} {\bf 414}, 279

\bibitem[\protect\astroncite{{Tinney} et~al.}{2005}]{tinney05}
{Tinney}, C.~G., {Burgasser}, A.~J., {Kirkpatrick}, J.~D., and {McElwain},
  M.~W.: 2005,
\newblock {\em \aj} {\bf 130}, 2326

\bibitem[\protect\astroncite{{Torres} et~al.}{2010}]{torres10}
{Torres}, G., {Andersen}, J., and {Gim{\'e}nez}, A.: 2010,
\newblock {\em \aapr} {\bf 18}, 67

\bibitem[\protect\astroncite{{Umbreit} et~al.}{2005}]{umbreit05}
{Umbreit}, S., {Burkert}, A., {Henning}, T., {Mikkola}, S., and {Spurzem}, R.:
  2005,
\newblock {\em \apj} {\bf 623}, 940

\bibitem[\protect\astroncite{{Valdivielso} et~al.}{2009}]{valdivielso09}
{Valdivielso}, L., {Mart{\'{\i}}n}, E.~L., {Bouy}, H., {Solano}, E., {Drew},
  J.~E., {Greimel}, R., {Guti{\'e}rrez}, R., {Unruh}, Y.~C., and {Vink}, J.~S.:
  2009,
\newblock {\em \aap} {\bf 497}, 973

\bibitem[\protect\astroncite{{van Rhijn}}{1936}]{vanrhijn36}
{van Rhijn}, P.~J.: 1936,
\newblock {\em Publications of the Kapteyn Astronomical Laboratory Groningen}
  {\bf 47}, 1

\bibitem[\protect\astroncite{{Vrba} et~al.}{2004}]{vrba04}
{Vrba}, F.~J., {Henden}, A.~A., {Luginbuhl}, C.~B., {Guetter}, H.~H., {Munn},
  J.~A., {Canzian}, B., {Burgasser}, A.~J., {Kirkpatrick}, J.~D., {Fan}, X.,
  {Geballe}, T.~R., {Golimowski}, D.~A., {Knapp}, G.~R., {Leggett}, S.~K.,
  {Schneider}, D.~P., and {Brinkmann}, J.: 2004,
\newblock {\em \aj} {\bf 127}, 2948

\bibitem[\protect\astroncite{{Whitworth} et~al.}{2007}]{whitworth07}
{Whitworth}, A., {Bate}, M.~R., {Nordlund}, {\AA}., {Reipurth}, B., and
  {Zinnecker}, H.: 2007,
\newblock {\em Protostars and Planets V} pp 459--476

\bibitem[\protect\astroncite{{Whitworth} et~al.}{2010}]{whitworth10}
{Whitworth}, A., {Stamatellos}, D., {Walch}, S., {Kaplan}, M., {Goodwin}, S.,
  {Hubber}, D., and {Parker}, R.: 2010,
\newblock in {R.~de Grijs \& J.~R.~D.~L{\'e}pine} (ed.), {\em IAU Symposium},
  Vol. 266 of {\em IAU Symposium}, pp 264--271

\bibitem[\protect\astroncite{{Whitworth} and {Stamatellos}}{2006}]{whitworth06}
{Whitworth}, A.~P. and {Stamatellos}, D.: 2006,
\newblock {\em \aap} {\bf 458}, 817

\bibitem[\protect\astroncite{{Wielen}}{1974}]{wielen74}
{Wielen}, R.: 1974,
\newblock {\em Highlights of Astronomy} {\bf 3}, 395

\bibitem[\protect\astroncite{{Xia} et~al.}{2008}]{xia08}
{Xia}, F., {Ren}, S., and {Fu}, Y.: 2008,
\newblock {\em \apss} {\bf 314}, 51

\bibitem[\protect\astroncite{{Zapatero Osorio} et~al.}{2000}]{zapatero00}
{Zapatero Osorio}, M.~R., {B{\'e}jar}, V.~J.~S., {Mart{\'{\i}}n}, E.~L.,
  {Rebolo}, R., {Barrado y Navascu{\'e}s}, D., {Bailer-Jones}, C.~A.~L., and
  {Mundt}, R.: 2000,
\newblock {\em Science} {\bf 290}, 103

\bibitem[\protect\astroncite{{Zapatero Osorio} et~al.}{2002}]{zapatero02}
{Zapatero Osorio}, M.~R., {B{\'e}jar}, V.~J.~S., {Pavlenko}, Y., {Rebolo}, R.,
  {Allende Prieto}, C., {Mart{\'{\i}}n}, E.~L., and {Garc{\'{\i}}a L{\'o}pez},
  R.~J.: 2002,
\newblock {\em \aap} {\bf 384}, 937

\bibitem[\protect\astroncite{{Zheng} et~al.}{2001}]{zheng01}
{Zheng}, Z., {Flynn}, C., {Gould}, A., {Bahcall}, J.~N., and {Salim}, S.: 2001,
\newblock {\em \apj} {\bf 555}, 393

\end{thebibliography}

\end{document}